\documentclass[%
preprint,
 amsmath,amssymb,
 aps,
prb,
]{revtex4-1}
\usepackage{graphicx}% Include figure files
\usepackage{dcolumn}% Align table columns on decimal point
\usepackage{bm}% bold math
\usepackage{amsmath}
\usepackage{amssymb}
\usepackage{comment}
\usepackage{setspace}
\usepackage{color}
\usepackage{float}

\newcommand{\cm}{cm$^{-1}$}
\newcommand{\cms}{cm$^{-1}$ }

\begin{document}

\title{Three-player polaritons: nonadiabatic fingerprints in an entangled atom-molecule-photon system}

\author{Tam\'as Szidarovszky}
    \email{tamas821@caesar.elte.hu}
    \affiliation{Laboratory of Molecular Structure and Dynamics, Institute of Chemistry, E\"otv\"os Lor\'and University and MTA-ELTE Complex Chemical Systems Research Group, 
	H-1117 Budapest, P\'azm\'any P\'eter s\'et\'any 1/A, Hungary}%

\author{G\'abor J. Hal\'asz}
	\affiliation{Department of Information Technology, University of Debrecen, P.O. Box 400, H-4002 Debrecen, Hungary}

\author{\'Agnes Vib\'ok}
    \email{vibok@phys.unideb.hu}
	\affiliation{Department of Theoretical Physics, University of Debrecen, P.O. Box 400, H-4002 Debrecen, Hungary and ELI-ALPS, ELI-HU Non-Profit Ltd., Dugonics t\'er 13, H-6720 Szeged, Hungary}	
\date{\today}% It is always \today, today,
             %  but any date may be explicitly specified

%\begin{tocentry}
%\includegraphics[width=1.0\textwidth]{TOC.png}
%\end{tocentry}

\begin{abstract}
A quantum system composed of a molecule and an atomic ensemble, confined in a microscopic cavity, is investigated theoretically. 
The indirect coupling between atoms and the molecule, realized by their interaction with the cavity radiation mode, leads to a coherent mixing of atomic and molecular states, and at strong enough cavity field strengths hybrid atom-molecule-photon polaritons are formed.
It is shown for the Na$_2$ molecule that by changing the cavity wavelength and the atomic transition frequency, the potential energy landscape of the polaritonic states and the corresponding spectrum could be changed significantly.
Moreover, an unforeseen intensity borrowing effect, which can be seen as a strong nonadiabatic fingerprint, is identified in the atomic transition peak, originating from the contamination of the atomic excited state with excited molecular rovibronic states.
\end{abstract}

\maketitle

\section{Introduction}
It is now understood that when molecules, viewed as quantum emitters, are placed inside an optical cavity and strongly interact with a confined radiation mode of the cavity, so-called polaritons are formed.\cite{cavity_Ebbesen_AccChemRes_2016,cavity_Feist_ACSphotonics_2018,Joel1a,theory_of_organic_cavities_Herrera_ACSphotonics_2018,Flick1,Ruggenthaler2018}
These are excited eigenstates, separated by the vacuum Rabi splitting, and they represent the interaction of the cavity radiation field with the molecular electronic excitations. 
As such, they are the quantum analogues of the well-known semiclassical molecular dressed states providing light-induced potentials.\cite{Bandrauk3}

Polaritonic chemistry has become an emerging research field, aimed at providing new tools for the fundamental investigation of %the 
light-matter interaction. 
Since the pioneering experimental work carried out the group of Ebbesen, in which they observed that strong light-matter coupling could modify %the 
chemical landscapes, \cite{cavity_exp_Hutchison_AngChem_2012} the field of ``molecular polaritons'' experienced much activity from both experimental \cite{Ebbesen3a,Ebbesen4a,Ebbesen5a,cavity_Zhong_AngChem_2016,Ebbesen7a,Ebbesen8a,Simon-Janos,Long1,cavity_Muallem_JPCL_2016,Edina1,Fleischer1} and theoretical \cite{cavity_Galego_PRX_2015,cavity_Galego_NatCommun_2016,markus1a,cavity_Kowalewski_JPCL_2016,cavity_Luk_JCTC_2017,gerit2a,gerit3a,joel2a,joel3a,joel4a,cavity_Herrera_PRL_2016,dark_vibronic_polaritons_Herrera_PRL_2017,cavity_Flick_PNAS_2017,cavityBO_Flick_JCTC_2017,cavity_MCTDH_Vendrell_CP_2018,Oriol2a,Fregoni2019,Schfer2019} research groups. 
Recent achievements on molecules strongly coupled to a cavity mode, such as that strong cavity-matter coupling can alter chemical reactivity,\cite{Ebbesen3a} provide long-range energy or charge transfer mechanisms,\cite{cavity_Zhong_AngChem_2016,Schfer2019} modify nonradiative relaxation pathways through collective effects,\cite{Oriol2a} and modify the optical response of molecules,\cite{dark_vibronic_polaritons_Herrera_PRL_2017,Tamas1a} support the relevance of such a new chemistry.
In most of the works published so far, many organic emitters were put into a cavity and the interest was mainly focused on the investigation of collective phenomena between the quantum radiation field and the molecular ensemble. 
In such theoretical descriptions, the molecules are usually treated with a reduced number of degrees of freedom or with some simplified models assuming two-level systems. \cite{cavity_Galego_PRX_2015}
However, it is worth studying single-molecule cavity %situations
interactions as well, since a more detailed study of individual objects may also provide meaningful results. \cite{markus1a,cavity_Kowalewski_JPCL_2016,Gabi6,Tamas1a,Agi2}

In the case of a single molecule, including several internal nuclear degrees of freedom gives rise to nonadiabatic dynamics. 
The radiation field can strongly mix the vibrational, rotational, and electronic degrees of freedom, thus creating light-induced avoided crossings (LIACs) or light-induced conical intersections (LICIs) between polaritonic potential-energy surfaces. \cite{Agi3}
Recent works discussed how the natural avoided crossings (already present in field-free systems) can be manipulated by placing the molecule into a nano-cavity and %treating 
influencing the ultrafast dynamics by means of a quantized radiation field. \cite{cavity_Kowalewski_JPCL_2016,Agi2}
Field-dressed rovibronic spectra of diatomics within the framework of cavity quantum electrodynamics (CQED) have also been investigated recently, \cite{Tamas1a} elucidating the significant contribution of molecular rotation to nonadiabatic effects in the spectrum.

In the present work, we investigate a complex quantum system consisting of three different types of entities; a cavity radiation mode, a molecule, and two-state atoms.
We examine how the hybrid atom-molecule-photon polaritonic energy surfaces change under different conditions, and how this affects the spectrum of each component.
We demonstrate how one component of the entangled system can manipulate the spectrum of another by means of indirect cavity coupling, thus resulting in the coherent mixing of atomic and molecular states and an unexpected strong nonadiabatic fingerprint in the atomic spectrum.

\section{Polaritonic landscape of a three-player system}
\subsection{Theoretical approach}
    Within the framework of QED and the electric dipole representation, the Hamiltonian of a molecule and $N_{\rm a}$ atoms interacting with a single cavity mode can be written as \cite{Cohen-Tannoudji}
    \begin{equation}
    \hat{H}_{{\rm tot}}=\hat{H}_{{\rm mol}}+\sum_{i=1}^{N_{\rm a}}\hslash\omega_a\hat{\sigma}_i^{\dagger}\hat{\sigma}_i+\hslash\omega _c\hat{a}^{\dagger}\hat{a}-\sqrt{\frac{\hslash\omega _c}{2\epsilon_{0}V}}\mathbf{\hat{d}\hat{e}}\left(\hat{a}^{\dagger}+\hat{a}\right),\label{eq:CavityHamiltonian_general}
    \end{equation}
    where $\hat{H}_{\rm mol}$ is the field-free molecular Hamiltonian, $\hat{\sigma_i}^{\dagger}$ and $\hat{\sigma_i}$ are excitation and deexcitation operators, respectively, for the $i$th atom, $\omega_a$ is the atomic transition frequency, $\hat{a}^{\dagger}$ and $\hat{a}$ are photon creation and annihilation operators, respectively, $\omega_c$ is the frequency
    of the cavity mode, $\hslash$ is Planck's constant divided by $2\pi$,
    $\epsilon_{0}$ is the electric constant, $V$ is the volume of the
    electromagnetic mode, $\mathbf{\hat{e}}$ is the polarization vector of the
    cavity mode, and $\mathbf{\hat{d}}=\mathbf{\hat{d}}^{\rm (mol)}+\mathbf{\hat{d}}^{\rm (a)}$ is the sum of the molecular and atomic dipole moments.
    
    It is useful to inspect the structure of the Hamiltonian of Eq. (\ref{eq:CavityHamiltonian_general}), when represented using $\vert \alpha \rangle \vert N \rangle \vert I \rangle$ product basis functions, where $\alpha$ and $I$ are the molecular electronic and atomic quantum numbers, respectively, and $N$ stands for the photon number. In our notation $I=0$ indicates that all atoms are in their ground state, and $I=n$ indicates that the $n$th atom is in an excited state, multiple atomic excitations are omitted in our model. The resulting Hamiltonian for a diatomic molecule reads (for simplicity we show the case when $\alpha=1$ or 2, $N=0$ or 1, and $I=0,1$ or 2, however, larger $N$ and $I$ values occur in the simulations below)
    
    \begin{equation}
    \hat{H}=
    \begin{bmatrix}\hat{H}_{{\rm m}} & \hat{A} & 0 & \hat{B} & 0 & \hat{B}\\
    \hat{A}^{\dagger} & \hat{H}_{{\rm m}}+\hslash\omega_c & \hat{B} & 0 & \hat{B} & 0 \\
    0 & \hat{B}^{\dagger} & \hat{H}_{{\rm m}}+\hslash\omega_a & \hat{A} & 0 & 0 \\
    \hat{B}^{\dagger} & 0 & \hat{A}^{\dagger} & \hat{H}_{{\rm m}}+\hslash\omega_a + \hslash\omega_c & 0 & 0 \\
    0 & \hat{B}^{\dagger} & 0 & 0 & \hat{H}_{{\rm m}}+\hslash\omega_a & \hat{A} \\
    \hat{B}^{\dagger} & 0 & 0 & 0 & \hat{A}^{\dagger} & \hat{H}_{{\rm m}}+\hslash\omega_a + \hslash\omega_c 
    \end{bmatrix},\label{eq:CavityHamiltonian}
    \end{equation}
    where 
    \begin{equation}
    \hat{H}_{{\rm m}}=\begin{bmatrix}\hat{T} & 0\\
    0 & \hat{T}
    \end{bmatrix}+\begin{bmatrix}V_{1}(R) & 0\\
    0 & V_{2}(R)
    \end{bmatrix},\label{eq:MolecularHamiltonian}
    \end{equation}
    \begin{equation}
    \hat{A}=\begin{bmatrix}g_{11}^{\rm (mol)}(R,\theta) & g_{12}^{\rm (mol)}(R,\theta)\\
    g_{21}^{\rm (mol)}(R,\theta) & g_{22}^{\rm (mol)}(R,\theta)
    \end{bmatrix} \quad {\rm and } \quad \hat{B}=\begin{bmatrix}g^{\rm (a)} & 0\\
    0 & g^{\rm (a)}
    \end{bmatrix},\label{eq:CavityAmx}
    \end{equation}
    with 
    \begin{equation}
    g_{ij}^{\rm (mol)}(R,\theta)=-\sqrt{\frac{\hslash\omega_c}{2\epsilon_{0}V}}d_{ij}^{\rm (mol)}(R){\rm cos}(\theta) \quad {\rm and} \quad g^{\rm (a)}=-\sqrt{\frac{\hslash\omega_c}{2\epsilon_{0}V}}d^{\rm (a)},\label{eq:quantized_A}
    \end{equation}
    where $R$ is the internuclear distance, $V_{i}(R)$ is the $i$th
    potential energy curve (PEC), $\hat{T}$ is the nuclear kinetic energy
    operator, $d_{ij}^{\rm (mol)}(R)$ is the transition dipole moment matrix element
    between the $i$th and $j$th molecular electronic states, $d^{\rm (a)}$ is the  transition dipole between the atomic ground and excited state, and $\theta$ is
    the angle between the electric field polarization vector and the molecular transition dipole vector, assumed to be parallel to the molecular axis. In Eq. (\ref{eq:CavityHamiltonian}) no rotating-wave approximation is applied, the accuracy of the model is only limited by the dipole approximation, the neglection of dissipative processes, and the limited number of electronic excitations considered for the molecule and atoms. Naturally, the maximum value of the photon number $N$ should be increased in practical applications until convergence is reached, resulting in a larger Hamiltonian and appropriate $\sqrt{N}$ factors for the $N>1$ $\hat{A}$ and $\hat{B}$ coupling terms.\cite{Cohen-Tannoudji}
    
    \begin{spacing}{1.0}
       \begin{figure}[H]
       \centering
        \includegraphics[width=0.49\textwidth]{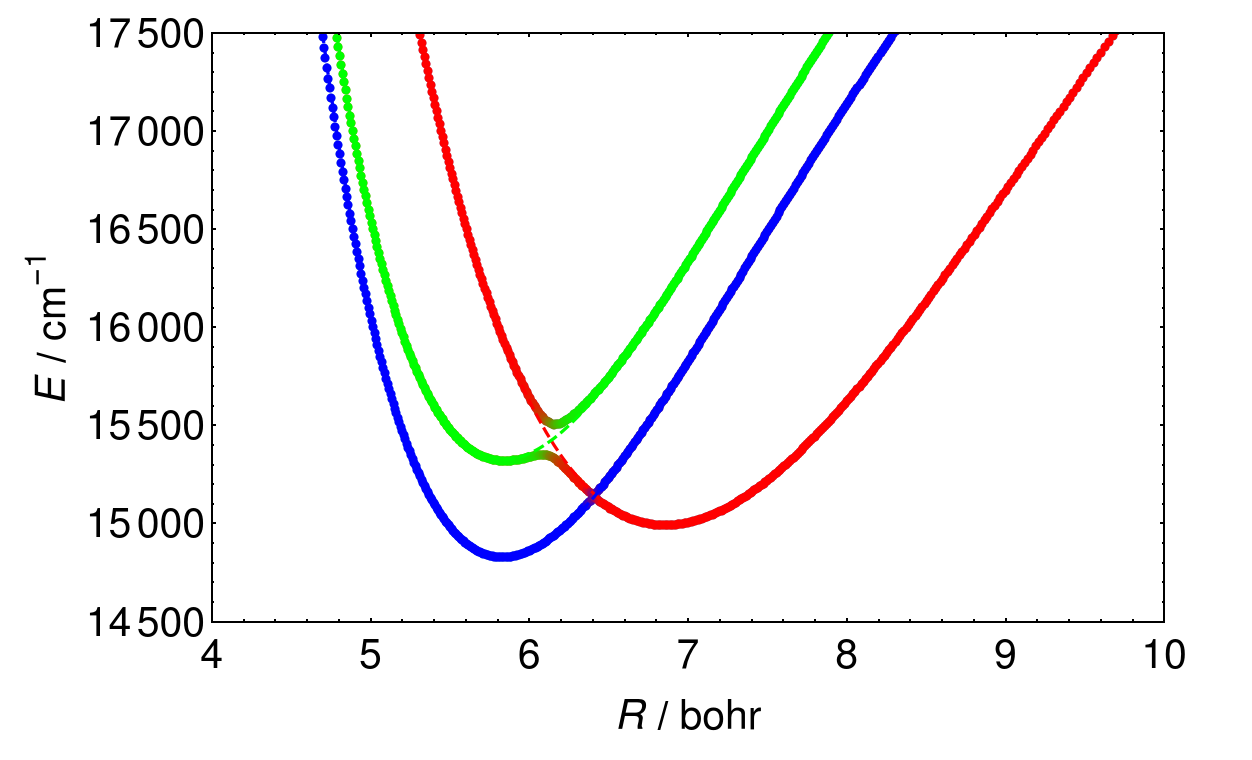}
        \includegraphics[width=0.45\textwidth]{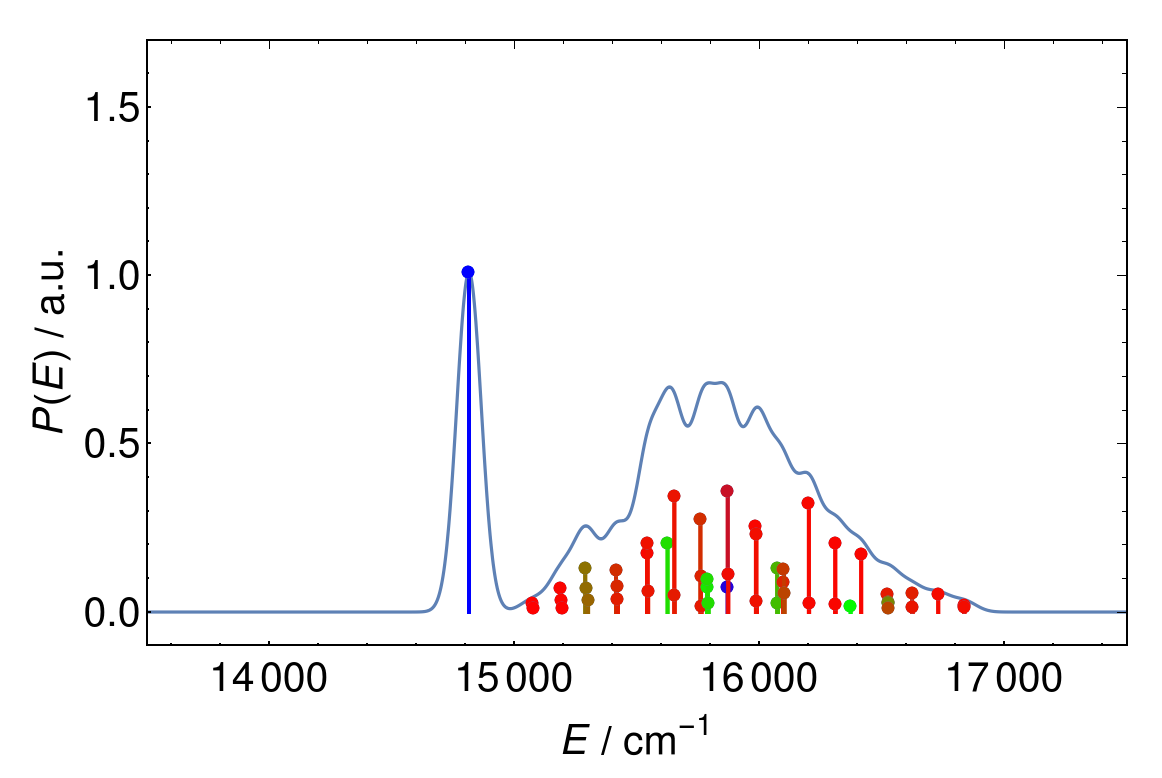}\\
        \includegraphics[width=0.49\textwidth]{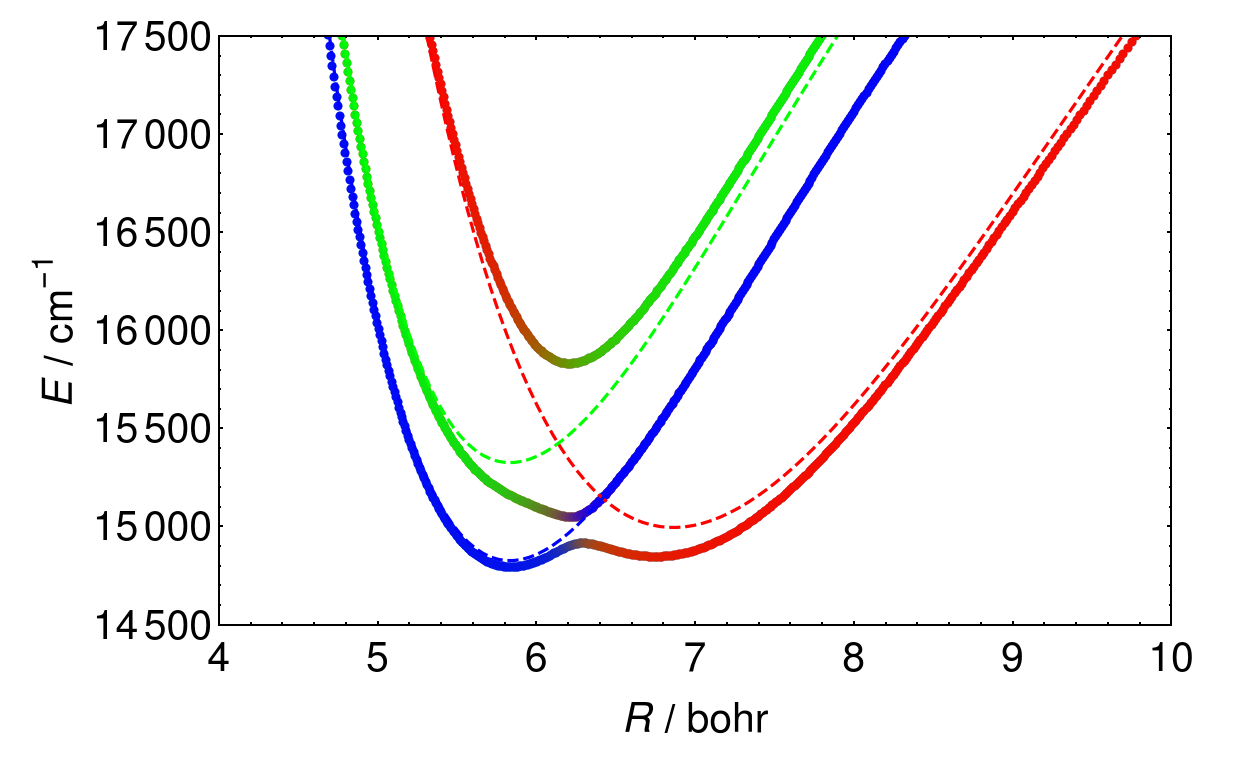}
        \includegraphics[width=0.46\textwidth]{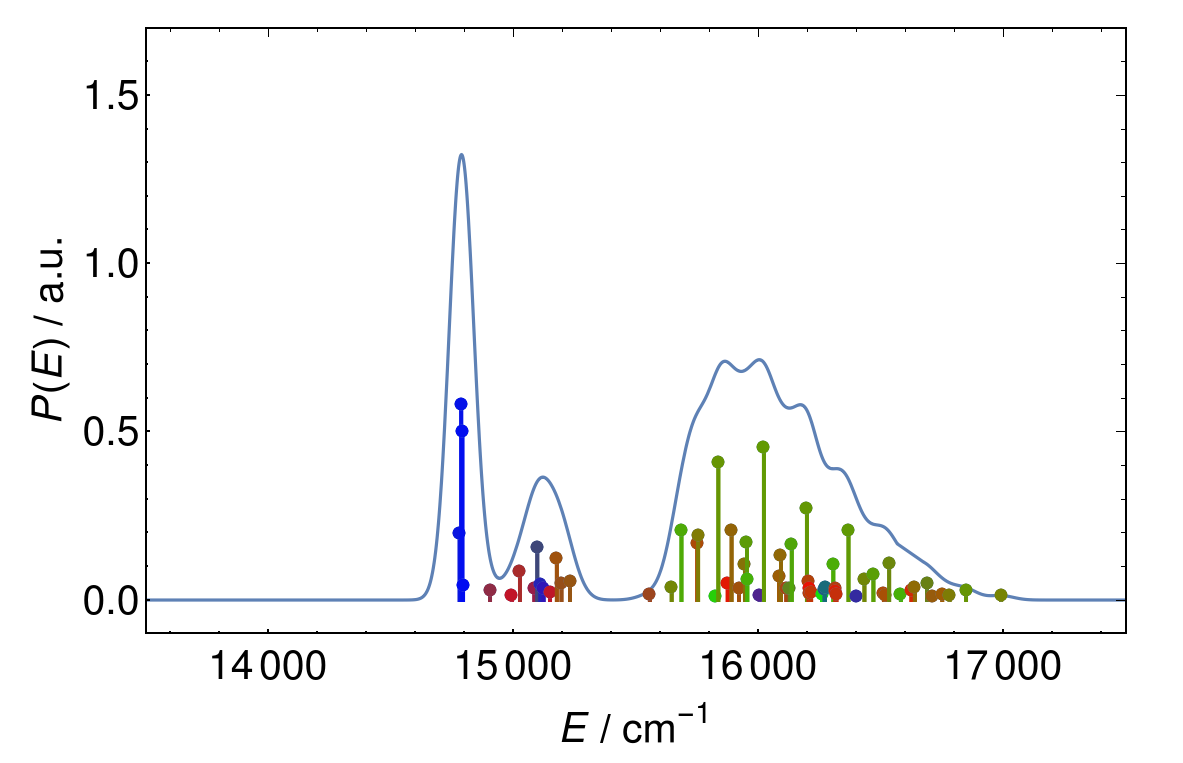}\\
        \includegraphics[width=0.49\textwidth]{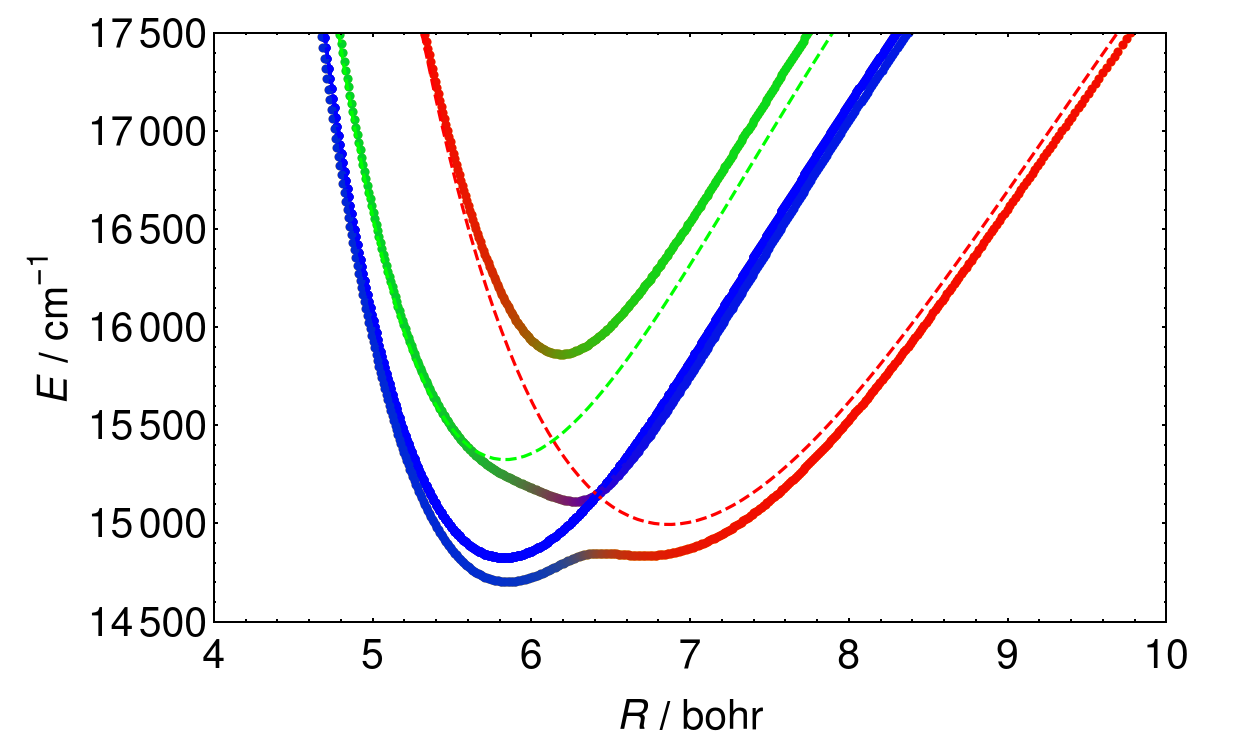}
        \includegraphics[width=0.46\textwidth]{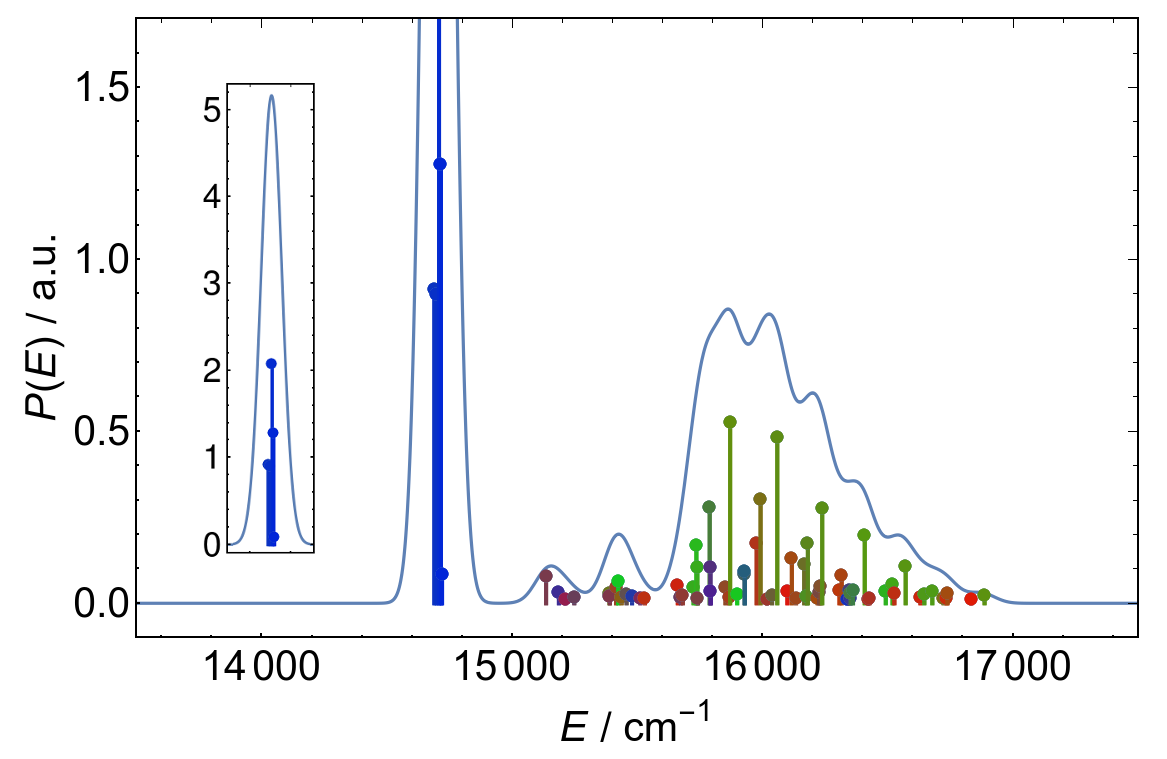}\\
        
        \caption{\footnotesize{
        \textbf{Left panels}: Diabatic (dashed lines) and adiabatic (solid lines) PECs of the Na$_{2}$ + $N_{\rm a}$ atom system confined in a cavity with radiation-field wavelength of $\lambda=653$ nm. The atomic excitation wavelength is 675 nm. Red, green, and blue colors represent the $\vert \alpha \rangle \vert N \rangle \vert I \rangle = \vert 2 \rangle \vert 0 \rangle \vert 0 \rangle, \vert 1 \rangle \vert 1 \rangle \vert 0 \rangle$ and $\vert 1 \rangle \vert 0 \rangle \vert 1 \rangle$ states, \textit{i.e.}, molecular electronic, photonic, and atomic excitations, respectively. The color of the adiabatic curves (polaritons) at each given $R$ value was generated by the mixture of red, green, and blue colors, with weights representing the weights of the different $\vert \alpha \rangle \vert N \rangle \vert I \rangle$ diabatic states within the given adiabatic state. The top and middle panels were generated with $N_{\rm a}=1$, and using a cavity-field strength of $\sqrt{\hslash\omega _c/2\epsilon_{0}V}=0.0001$ au and 0.0005 au, respectively, while the bottom panel was generated with $N_{\rm a}=5$ and $\sqrt{\hslash\omega _c/2\epsilon_{0}V}=0.0005$ au.
        \textbf{Right panels}: Absorption spectra corresponding to the PECs shown in the left panels. Red, green, and blue colors represent the $\vert \alpha v J \rangle \vert N \rangle \vert I \rangle = \vert 2 v J \rangle \vert 0 \rangle \vert 0 \rangle, \vert 1 v J \rangle \vert 1 \rangle \vert 0 \rangle$ and $\vert 1 v J \rangle \vert 0 \rangle \vert 1 \rangle$-type states, \textit{i.e.}, molecular electronic, photonic, and atomic excitations, respectively. Note that molecular rovibrational excitation can occur in all three cases. The color of the individual peaks was generated by the mixture of red, green, and blue colors, with weights representing the weights of the different $\vert \alpha v J \rangle \vert N \rangle \vert I \rangle$ states within the dressed state to which the transition occurs. The inset in the lower right panel shows the atomic transitions around 14700 \cms in their full height.
        }}
        \label{fig:PECs_and_spectra_atr_675}
    \end{figure}
    \end{spacing}

\subsection{Numerical results}

    Based on the Hamiltonian of Eq. (\ref{eq:CavityHamiltonian_general}), the left panels of Figure \ref{fig:PECs_and_spectra_atr_675} show the potential energy landscape of a Na$_2$ molecule and two-state atoms interacting with a single cavity radiation mode. For Na$_{2}$ the $V_{1}(R)$ and $V_{2}(R)$ PECs correspond to the ${\rm X}^{1}\Sigma{\rm _{g}^{+}}$ and the ${\rm A}^{1}\Sigma{\rm _{u}^{+}}$ electronic states, respectively, taken from Ref. \citenum{Na2_PEC_Magnier_JCP_1993}. For Na$_2$ $d_{11}^{\rm (mol)}(R)=d_{22}^{\rm (mol)}(R)=0$ and the $d_{12}^{\rm (mol)}(R)$ transition dipole was taken from Ref. \citenum{Na2_TDM_Zemke_JMS_1981}, while $d^{\rm (a)}=1$ au was used for the atoms.
    Fig. \ref{fig:PECs_and_spectra_atr_675} demonstrates that the coupling with the radiation field leads to the formation of three ``bright'' polaritonic surfaces, and as shown in the bottom left panel, $N_{\rm a}-1$ ``dark states''. Although these ``dark states'' can affect excited state dynamics,\cite{Oriol2a} they usually have no significant contribution to the spectrum.\cite{Houdr1996,Joel1a} With increasing atom number the effective coupling with the field increases and the polaritonic surfaces become more pronounced.

% Fig 1 original place 

\begin{figure}[h!]
    \centering
        \includegraphics[width=0.49\textwidth]{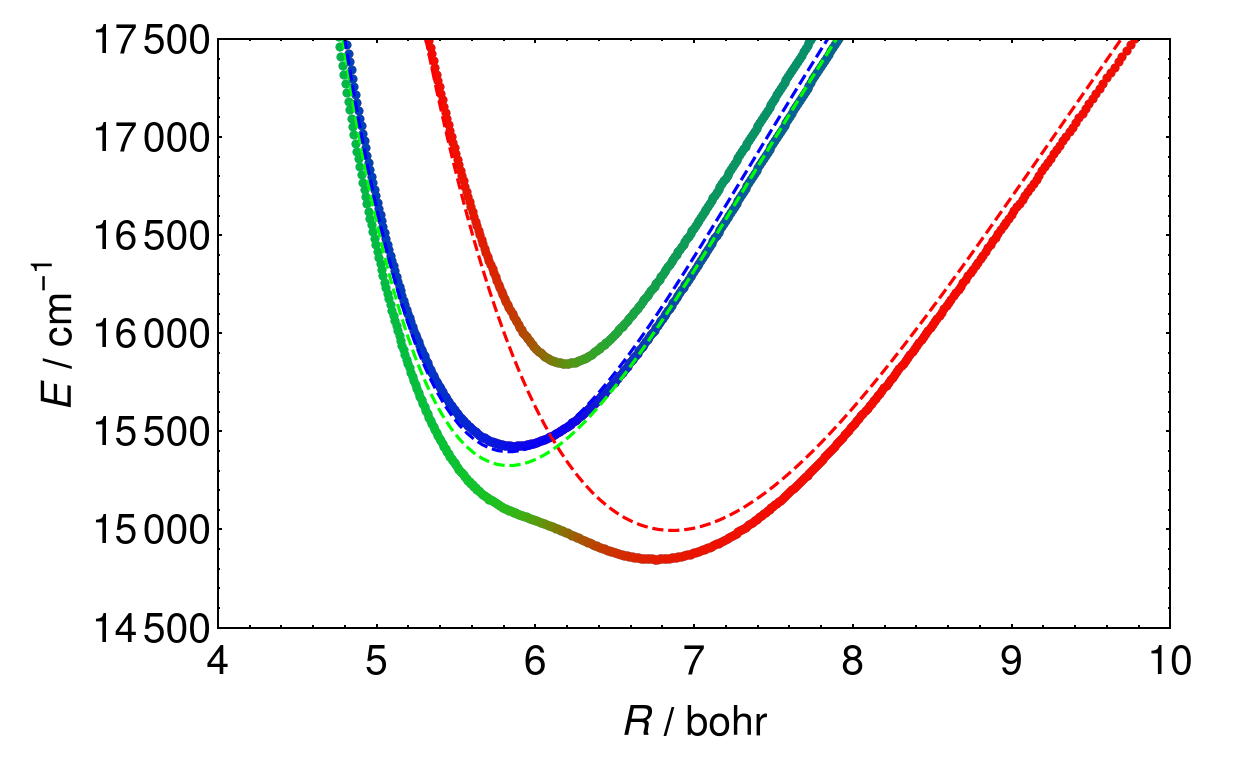}
        \includegraphics[width=0.45\textwidth]{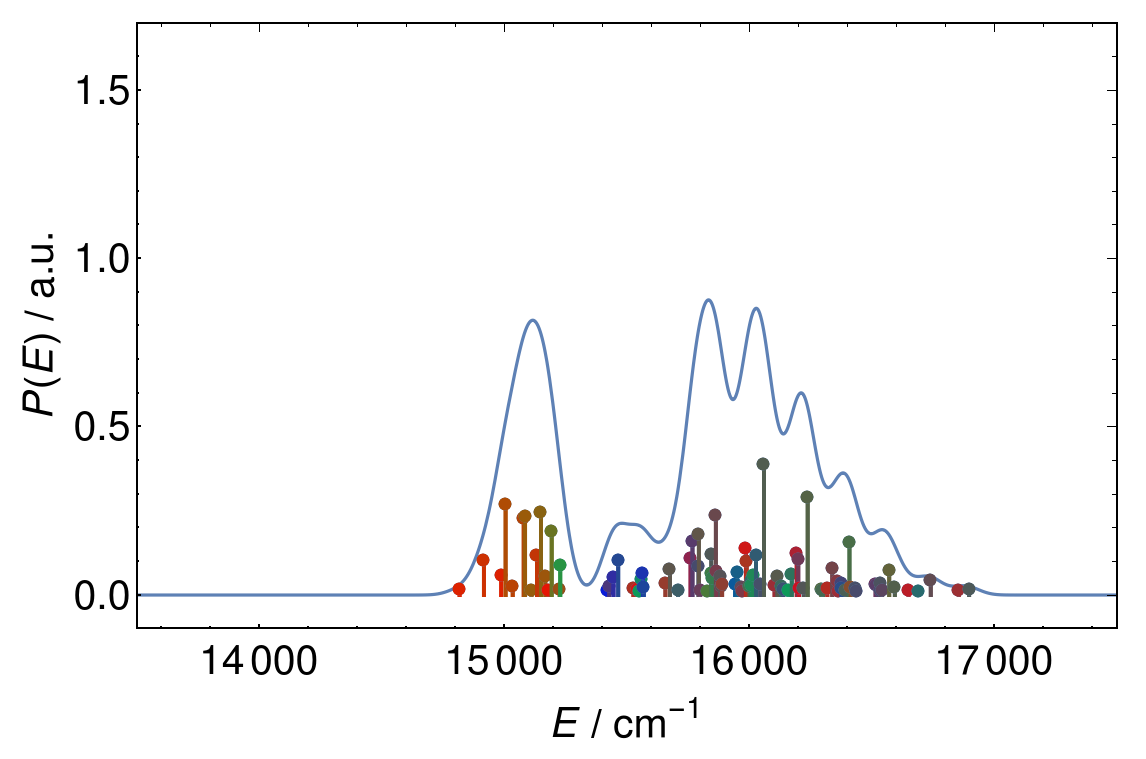}
        \includegraphics[width=0.49\textwidth]{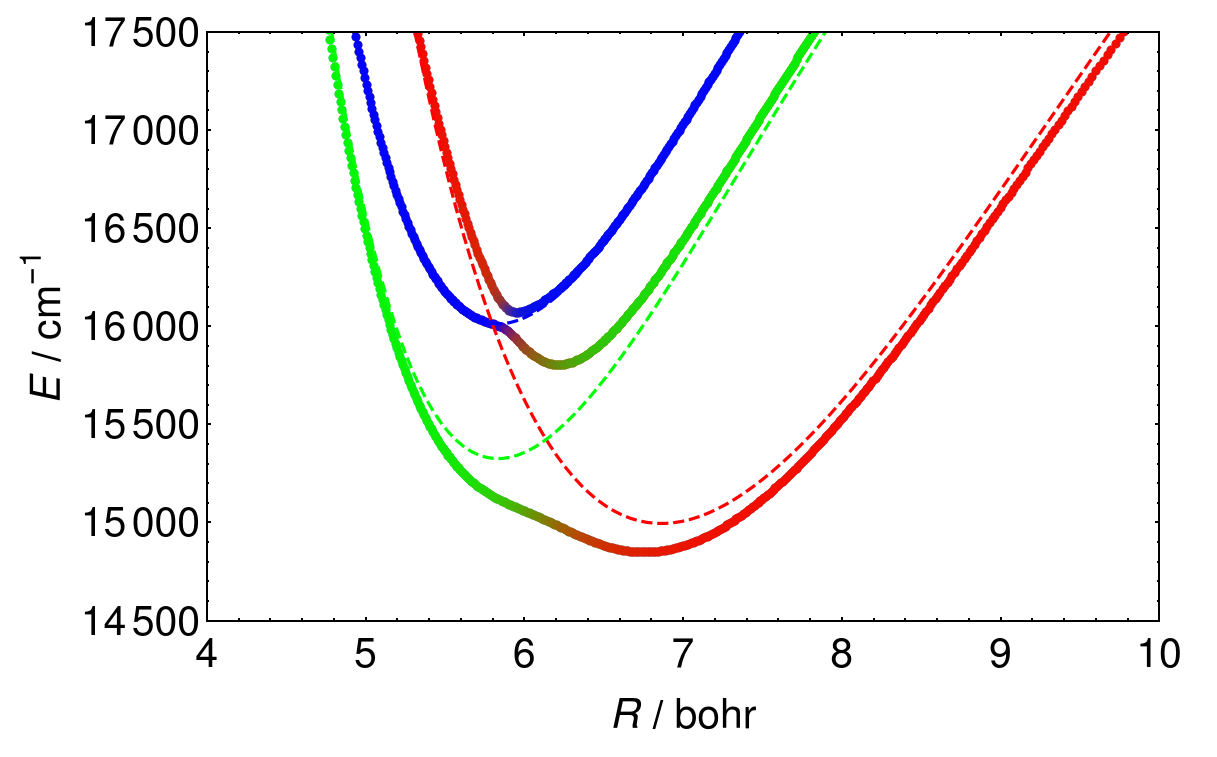}
        \includegraphics[width=0.46\textwidth]{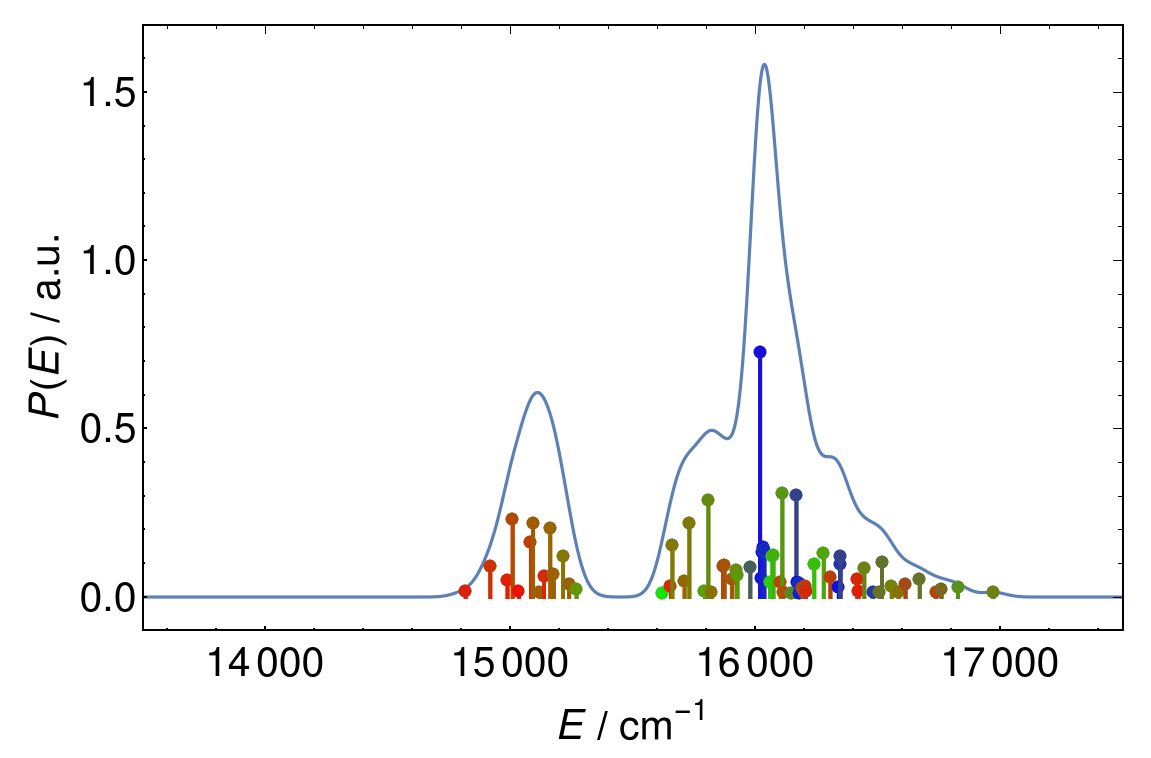}
        \includegraphics[width=0.49\textwidth]{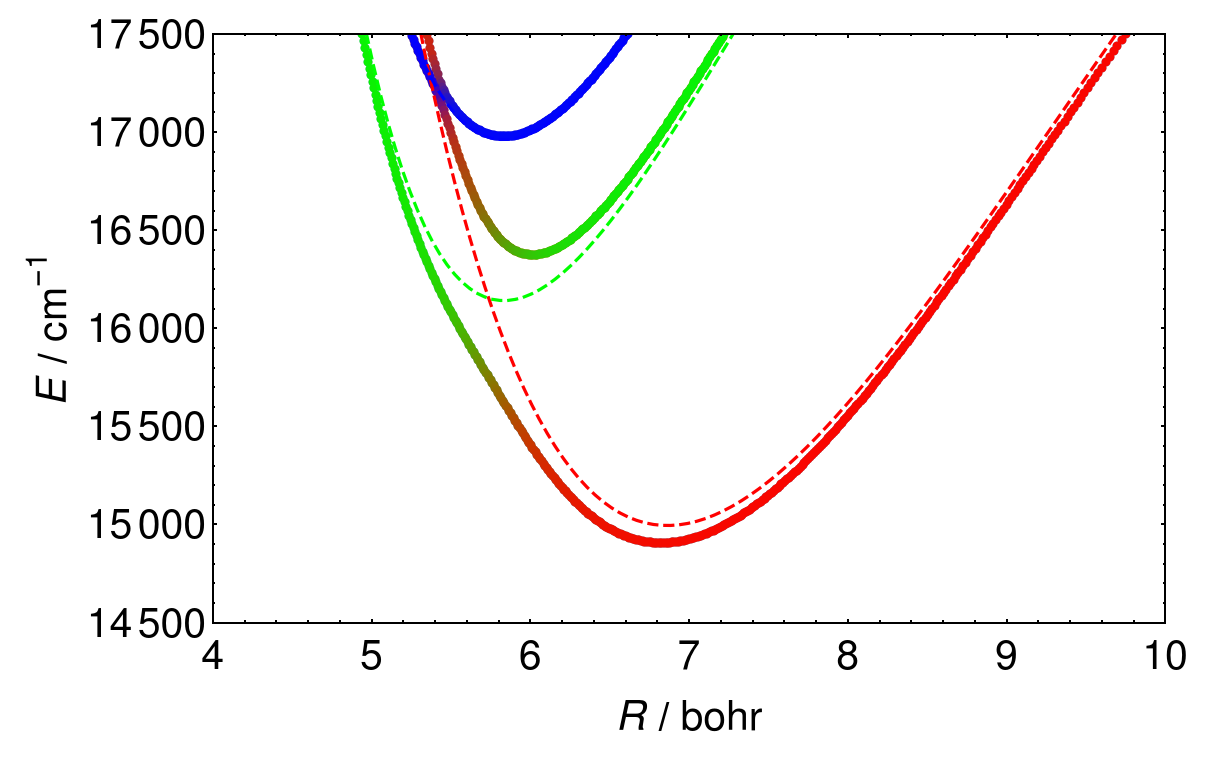}
        \includegraphics[width=0.46\textwidth]{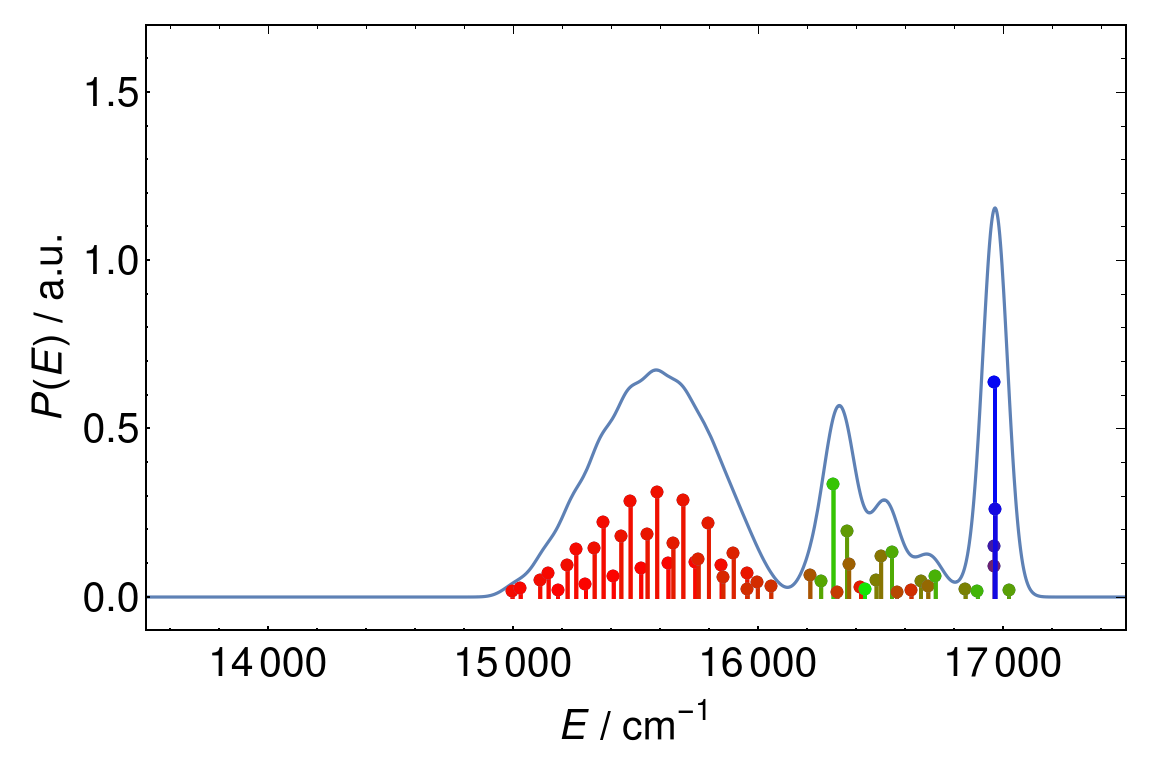}
        
        \caption{Same as in Fig. \ref{fig:PECs_and_spectra_atr_675}, but with a cavity-field strength of $\sqrt{\hslash\omega _c/2\epsilon_{0}V}=0.0005$ au for all panels, and with the following radiation-field wavelengths $\lambda_c$ and atomic excitation wavelengths $\lambda_a$. Upper panels: $\lambda_c=653$ nm, $\lambda_a=650$ nm, middle panels: $\lambda_c=653$ nm, $\lambda_a=625$ nm, and bottom panels: $\lambda_c=620$ nm, $\lambda_a=590$ nm.}
        \label{fig:PECs_and_spectra_mixed}
    \end{figure}

    The left panels of Figure \ref{fig:PECs_and_spectra_mixed} show the polaritonic states formed for different atomic and photonic excitation energies. Note that the 590 nm wavelength for atomic excitation shown in the lower left panel of Fig. \ref{fig:PECs_and_spectra_mixed} stands for the ${^2}{\rm P} \leftarrow {^2}{\rm S}$ transition of Na atoms. It is clear from Figs. \ref{fig:PECs_and_spectra_atr_675} and \ref{fig:PECs_and_spectra_mixed} that the presence of atoms, and their indirect coupling with molecules through the radiation mode, can considerably change the polaritonic landscape of molecules confined in cavities, providing an additional degree of freedom to modify and manipulate the excited state properties and dynamics of molecules.
    
\section{Absorption spectrum}
\subsection{Theoretical approach}
    Now we turn to the weak-field absorption spectrum of the composite system. Similar to previous works,\cite{Tamas1a,LICI_in_spectrum_Szidarovszky_JPCL_2018} the spectrum is determined by first computing the field-dressed states, \textit{i.e.}, the eigenstates of the full ``molecule + atoms + radiation field'' system, and then computing the dipole transition amplitudes between the field-dressed states with respect to a probe pulse. 
    The probe pulse is assumed to be weak, inducing primarily one-photon processes, implying that the standard approach \cite{BunkerJensen} of using first-order time-dependent perturbation theory to compute the transition amplitudes should be adequate.
    
    We determine the field-dressed states of the Hamiltonian in Eq. (\ref{eq:CavityHamiltonian_general}) by diagonlizing its matrix representation obtained with $\vert \alpha v J \rangle \vert N \rangle \vert I \rangle$ product basis functions, where $v$ and $J$ are the molecular vibrational and rotational quantum numbers, respectively. 
    The field-free molecular rovibronic states $\vert \alpha v J \rangle$ are computed using 200 discrete variable representation (DVR) spherical-DVR basis functions\cite{D2FOPI_Szidarovszky_PCCP_2010} for each PEC with the related grid points placed in the internuclear coordinate range $(0,10)$ bohr. All states with $J<16$ and an energy not exceeding the zero point energy of the respective PEC by more than 2000 cm$^{-1}$ were included into the product basis. The maximum photon number in the cavity mode was set to two.
    
    The \textit{i}th field-dressed state can then be expressed as 
    \begin{equation}
        \vert\Psi_{i}^{{\rm FD}}\rangle=\sum_{\alpha,v,J,N,I}C_{i,\alpha vJNI}\vert\alpha vJ\rangle\vert N\rangle \vert I\rangle,\label{eq:FieldDressedStates}
    \end{equation}
    where the $C_{i,\alpha vJNI}$ coefficients are the eigenvector components obtained from the diagonalization. 
    Using first-order time-dependent perturbation theory, the transition amplitude between two field-dressed states, induced by the weak probe pulse, is proportional to\cite{Jaynes_Cummings_1963,Cohen-Tannoudji}
    \begin{equation}
    \langle\Psi_{i}^{{\rm FD}}\vert\langle M\vert\mathbf{\hat{d}\hat{E}}\vert M'\rangle\vert\Psi_{j}^{{\rm FD}}\rangle=\Big( \langle\Psi_{i}^{{\rm FD}}\vert \hat{d}^{\rm (mol)}{\rm cos}(\theta)\vert\Psi_{j}^{{\rm FD}}\rangle+\langle\Psi_{i}^{{\rm FD}}\vert \hat{d}^{\rm (a)}\vert\Psi_{j}^{{\rm FD}}\rangle \Big) \langle M\vert\hat{E}\vert M'\rangle.\label{eq:transition_amplitude_general}
    \end{equation}
    In Eq. (\ref{eq:transition_amplitude_general}), $\hat{E}$ is the electric field
    operator of the weak probe pulse, assumed to have a polarization axis identical to
    that of the cavity mode, leading to $M=M'\pm1$ for the probe pulse. Thus, Eq. (\ref{eq:transition_amplitude_general}) accounts for single-photon absorption or stimulated emission. When the field-dressed states are expanded according to Eq. (\ref{eq:FieldDressedStates}), many terms in Eq. (\ref{eq:transition_amplitude_general}) become zero due to the selection rules of the atomic and molecular transitions involved. For a homonuclear diatomic molecule, such as Na$_2$, the $T_{j \leftarrow i}$ transition probability is proportional to
    \begin{equation}
        \bigg| \sum_{\alpha v J N I} \bigg( \sum_{\alpha' v' J'} C^*_{i,\alpha vJNI}C_{j,\alpha' v'J'NI} \langle v \vert d_{\alpha \alpha'}^{\rm (mol)}(R)\vert v' \rangle \langle J \vert {\rm cos}(\theta) \vert J' \rangle + \sum_{I'\neq I} C^*_{i,\alpha vJNI}C_{j,\alpha vJNI'} d^{\rm (a)} \bigg) \bigg|^2 
        \label{eq:transition_probability}
    \end{equation}
    Eq. (\ref{eq:transition_probability}) demonstrates that in the general field-dressed case the spectral peaks cannot be evaluated as the sum of field-dressed atomic and field-dressed molecular peaks, because of the interference between atomic and molecular transitions. Thus, fingerprints of the coherent mixing of atomic and molecular states can appear in the spectrum of the composite system. 
    Naturally, in the limit of the cavity field strength going to zero (and simultaneously the molecule-atom interaction going to zero) the transitions should reflect a spectrum composed as the sum of the molecular spectrum and the atomic spectrum. 
    However, when molecular and atomic couplings with the cavity are strong enough for polaritonic states to be formed, the spectrum should reflect transitions corresponding to field-dressed states, which are a coherent mixture of states containing atomic, molecular, and photonic excitations. 

\subsection{Results and discussion}

    It is well known that the interaction of atoms with a (near) resonant electromagnetic field leads to a Rabi splitting of the energy levels in the excited state manifold. This mechanism leads to splittings in the spectral peaks of the light-dressed system, often referred to in spectroscopy as Autler--Townes splitting.\cite{AutlerTownes_original}
    When $N_{\rm a}$ atoms simultaneously interact with the same mode of an electromagnetic field, a pair of splitted states appear, reflecting an effective coupling strength increased by $\sqrt{N_{\rm a}}$, accompanied by $N_{\rm a}-1$ dark states with no shift appearing in their energy with respect to the field-free atomic states.\cite{Houdr1996} Depending on the homogeneity of the atom-field interactions, with inhomogeneity caused either by inhomogeneities in the field or the atomic transition frequencies, the energy level structure and the absorption spectrum could become more complex, as detailed in Ref. \citenum{Houdr1996}.
    As for molecules interacting with a cavity radiation mode, Autler--Townes-type splittings, as well as ``intensity borrowing'' effects\cite{Cederbaum_multimode,BunkerJensen} and the formation of polaritonic potential energy surfaces, are reflected in the spectrum.\cite{Tamas1a}   
    The right panels of Fig. \ref{fig:PECs_and_spectra_atr_675} show the light-dressed spectra of an Na$_2$ molecule and two-state atoms interacting with a single cavity mode.
    All plotted peaks represent transitions from the ground state of the entangled ``atom+molecule+cavity'' system, primarily composed of $\vert 1 0 0\rangle \vert 0 \rangle \vert 0 \rangle$, to higher-lying field-dressed states, formed as a quantum superposition of molecular, atomic and photonic excited states. The composition of each mixed state is indicated by the color of its respective transition line, see details in the footnote of Fig. \ref{fig:PECs_and_spectra_atr_675}.
    In order to help identify overall intensity changes in the spectra, in addition to the peak splittings, an envelope for each spectrum was generated by convolving the peaks with a Gaussian function having a standard deviation of $\sigma = 50$ \cm.
    
    The upper right panel of Fig. \ref{fig:PECs_and_spectra_atr_675} demonstrates that at lower coupling strengths a light-dressed spectrum of the molecule, and a well defined, separate atomic transition peak at 14800 \cms can be observed. 
    The peaks between $15000-17000$ \cms mostly reflect transitions similar to the field-free molecular transitions $\vert 2 v 1 \rangle \leftarrow \vert 1 0 0\rangle$, however, the presence of the coupling between the molecule and the cavity field is also visible in peak splittings and in the colors of some transitions deviating from red.
    With increasing coupling strength the molecular, atomic, and photonic states become entangled and three bright polaritonic states are formed, which is reflected in the three groups of peaks in the spectrum (see middle panels of Fig. \ref{fig:PECs_and_spectra_atr_675}). 
    The splitting of the atomic peak at 14800 \cms in the middle right panel of Fig. \ref{fig:PECs_and_spectra_atr_675} as well as the appearance of purple colored peaks near 15100 \cms indicates that a considerable mixing of the atomic and molecular states occurred.
    Increasing the atom number to $N_{\rm a}$ while keeping the cavity field strength fixed leads to (1) an $\sqrt{N_{\rm a}}$ times increased effective atom-cavity coupling, resulting in a shift in the atomic transition frequency, (2) an $N_{\rm a}$ times increased atomic line strengths, and (3) the appearance of $N_{\rm a}-1$ dark states, see lower left panel of Fig. \ref{fig:PECs_and_spectra_atr_675}. Our numerical results show, as expected, that these dark states have no significant contribution to the spectrum (see lower right panel of Fig. \ref{fig:PECs_and_spectra_atr_675}).
    
    A striking feature in Fig. \ref{fig:PECs_and_spectra_atr_675} is that the roughly 1.3 overall peak intensity of the split atomic transition in the middle right panel is significantly different from the 1.0 intensity of the unsplit atomic peak of the upper right panel. This indicates an intensity borrowing effect in the atomic spectrum, \textit{i.e.}, the intensity of the atomic transition peak is modified, due to the contamination of the atomic excited state with molecular excited rovibronic states. This striking new phenomenon in the atomic spectrum is caused by the indirect coupling with the molecule, attributed to the radiation mode of the cavity.

    The right panels of Figure \ref{fig:PECs_and_spectra_mixed} show the absorption spectra obtained for the polaritonic states presented in the left panels of Fig. \ref{fig:PECs_and_spectra_mixed}. The position and the composition of the dressed states, represented by the color of their transition peaks, reflect well the corresponding polaritonic landscapes.
    For example, peaks with various colors between $15600-17000$ \cms in the upper right panel of Fig. \ref{fig:PECs_and_spectra_mixed} show the fact that in that energy region the polaritonic states (see upper left panel of Fig. \ref{fig:PECs_and_spectra_mixed}) are composed of a strong mixture of molecular, atomic, as well as photonic excited states, represented by basis functions of the form $\vert 2 v J\rangle \vert 0 \rangle \vert 0 \rangle$, $\vert 1 v J\rangle \vert 0 \rangle \vert I \neq 0 \rangle$, and $\vert 1 v J\rangle \vert 1 \rangle \vert 0 \rangle$, respectively.
    On the other hand, the red peaks between $15000-16000$ \cms in the lower right panel of Fig. \ref{fig:PECs_and_spectra_mixed} indicate that the corresponding polaritonic state includes mostly molecular excitation.
    Fig. \ref{fig:PECs_and_spectra_mixed} demonstrates that the response of molecules (atoms) to light can be significantly modified and manipulated by the presence of atoms (molecules), and their simultaneous interaction with a cavity radiation mode.
\section{Summary}
    To summarize, we demonstrated that when the composite system of a molecule and an atomic ensemble is confined in a microscopic cavity, then (1) three bright and $N_{\rm a}-1$ dark polaritonic states are formed, given that the cavity field strength is strong enough. 
    (2) The indirect coupling between atoms and molecule, due to their interaction with the cavity radiation mode, leads to a coherent mixing of atomic and molecular states, and the general field-dressed states of the excited state manifold carry signatures of atomic, molecular, as well as photonic excitations.
    (3) The absorption spectrum of the confined entangled system reflects well the polaritonic landscape, and an intensity borrowing effect could be identified in the atomic transition peak, originating from the contamination of the atomic excited state with excited molecular rovibronic states. 
    (4) It was shown that by changing the cavity wavelength and the atomic transition frequency, the potential energy landscape of the polaritonic states and the corresponding spectrum could be altered significantly. 
    This shows, that by adding a second type of entity to a quantum system confined in a microscopic cavity, the dynamics of the system and its response to light can be significantly modified and manipulated. This extra degree of freedom implies new possible avenues in polaritonic chemistry.

\section{Acknowledgement}

    This research was supported by the EU-funded Hungarian grant EFOP-3.6.2-16-2017-00005. The authors are grateful to NKFIH for support (Grant No. PD124623 and K119658).

\bibliography{Na2_and_atoms_in_cavity}

%merlin.mbs apsrev4-1.bst 2010-07-25 4.21a (PWD, AO, DPC) hacked
%Control: key (0)
%Control: author (8) initials jnrlst
%Control: editor formatted (1) identically to author
%Control: production of article title (-1) disabled
%Control: page (0) single
%Control: year (1) truncated
%Control: production of eprint (0) enabled
\begin{thebibliography}{51}%
\makeatletter
\providecommand \@ifxundefined [1]{%
 \@ifx{#1\undefined}
}%
\providecommand \@ifnum [1]{%
 \ifnum #1\expandafter \@firstoftwo
 \else \expandafter \@secondoftwo
 \fi
}%
\providecommand \@ifx [1]{%
 \ifx #1\expandafter \@firstoftwo
 \else \expandafter \@secondoftwo
 \fi
}%
\providecommand \natexlab [1]{#1}%
\providecommand \enquote  [1]{``#1''}%
\providecommand \bibnamefont  [1]{#1}%
\providecommand \bibfnamefont [1]{#1}%
\providecommand \citenamefont [1]{#1}%
\providecommand \href@noop [0]{\@secondoftwo}%
\providecommand \href [0]{\begingroup \@sanitize@url \@href}%
\providecommand \@href[1]{\@@startlink{#1}\@@href}%
\providecommand \@@href[1]{\endgroup#1\@@endlink}%
\providecommand \@sanitize@url [0]{\catcode `\\12\catcode `\$12\catcode
  `\&12\catcode `\#12\catcode `\^12\catcode `\_12\catcode `\%12\relax}%
\providecommand \@@startlink[1]{}%
\providecommand \@@endlink[0]{}%
\providecommand \url  [0]{\begingroup\@sanitize@url \@url }%
\providecommand \@url [1]{\endgroup\@href {#1}{\urlprefix }}%
\providecommand \urlprefix  [0]{URL }%
\providecommand \Eprint [0]{\href }%
\providecommand \doibase [0]{http://dx.doi.org/}%
\providecommand \selectlanguage [0]{\@gobble}%
\providecommand \bibinfo  [0]{\@secondoftwo}%
\providecommand \bibfield  [0]{\@secondoftwo}%
\providecommand \translation [1]{[#1]}%
\providecommand \BibitemOpen [0]{}%
\providecommand \bibitemStop [0]{}%
\providecommand \bibitemNoStop [0]{.\EOS\space}%
\providecommand \EOS [0]{\spacefactor3000\relax}%
\providecommand \BibitemShut  [1]{\csname bibitem#1\endcsname}%
\let\auto@bib@innerbib\@empty
%</preamble>
\bibitem [{\citenamefont {Ebbesen}(2016)}]{cavity_Ebbesen_AccChemRes_2016}%
  \BibitemOpen
  \bibfield  {author} {\bibinfo {author} {\bibfnamefont {T.~W.}\ \bibnamefont
  {Ebbesen}},\ }\href {\doibase 10.1021/acs.accounts.6b00295} {\bibfield
  {journal} {\bibinfo  {journal} {Acc. Chem. Res.}\ }\textbf {\bibinfo {volume}
  {49}},\ \bibinfo {pages} {2403} (\bibinfo {year} {2016})}\BibitemShut
  {NoStop}%
\bibitem [{\citenamefont {Feist}\ \emph {et~al.}(2018)\citenamefont {Feist},
  \citenamefont {Galego},\ and\ \citenamefont
  {Garcia-Vidal}}]{cavity_Feist_ACSphotonics_2018}%
  \BibitemOpen
  \bibfield  {author} {\bibinfo {author} {\bibfnamefont {J.}~\bibnamefont
  {Feist}}, \bibinfo {author} {\bibfnamefont {J.}~\bibnamefont {Galego}}, \
  and\ \bibinfo {author} {\bibfnamefont {F.~J.}\ \bibnamefont {Garcia-Vidal}},\
  }\href {\doibase 10.1021/acsphotonics.7b00680} {\bibfield  {journal}
  {\bibinfo  {journal} {ACS Photonics}\ }\textbf {\bibinfo {volume} {5}},\
  \bibinfo {pages} {205} (\bibinfo {year} {2018})}\BibitemShut {NoStop}%
\bibitem [{\citenamefont {Ribeiro}\ \emph {et~al.}(2018)\citenamefont
  {Ribeiro}, \citenamefont {Mart{\'{\i}}nez-Mart{\'{\i}}nez}, \citenamefont
  {Du}, \citenamefont {Campos-Gonzalez-Angulo},\ and\ \citenamefont
  {Yuen-Zhou}}]{Joel1a}%
  \BibitemOpen
  \bibfield  {author} {\bibinfo {author} {\bibfnamefont {R.~F.}\ \bibnamefont
  {Ribeiro}}, \bibinfo {author} {\bibfnamefont {L.~A.}\ \bibnamefont
  {Mart{\'{\i}}nez-Mart{\'{\i}}nez}}, \bibinfo {author} {\bibfnamefont
  {M.}~\bibnamefont {Du}}, \bibinfo {author} {\bibfnamefont {J.}~\bibnamefont
  {Campos-Gonzalez-Angulo}}, \ and\ \bibinfo {author} {\bibfnamefont
  {J.}~\bibnamefont {Yuen-Zhou}},\ }\href {\doibase 10.1039/c8sc01043a}
  {\bibfield  {journal} {\bibinfo  {journal} {Chem. Sci.}\ }\textbf {\bibinfo
  {volume} {9}},\ \bibinfo {pages} {6325} (\bibinfo {year} {2018})}\BibitemShut
  {NoStop}%
\bibitem [{\citenamefont {Herrera}\ and\ \citenamefont
  {Spano}(2018)}]{theory_of_organic_cavities_Herrera_ACSphotonics_2018}%
  \BibitemOpen
  \bibfield  {author} {\bibinfo {author} {\bibfnamefont {F.}~\bibnamefont
  {Herrera}}\ and\ \bibinfo {author} {\bibfnamefont {F.~C.}\ \bibnamefont
  {Spano}},\ }\href {\doibase 10.1021/acsphotonics.7b00728} {\bibfield
  {journal} {\bibinfo  {journal} {ACS Photonics}\ }\textbf {\bibinfo {volume}
  {5}},\ \bibinfo {pages} {65} (\bibinfo {year} {2018})}\BibitemShut {NoStop}%
\bibitem [{\citenamefont {Flick}\ \emph {et~al.}(2018)\citenamefont {Flick},
  \citenamefont {Rivera},\ and\ \citenamefont {Narang}}]{Flick1}%
  \BibitemOpen
  \bibfield  {author} {\bibinfo {author} {\bibfnamefont {J.}~\bibnamefont
  {Flick}}, \bibinfo {author} {\bibfnamefont {N.}~\bibnamefont {Rivera}}, \
  and\ \bibinfo {author} {\bibfnamefont {P.}~\bibnamefont {Narang}},\ }\href
  {\doibase 10.1515/nanoph-2018-0067} {\bibfield  {journal} {\bibinfo
  {journal} {Nanophotonics}\ }\textbf {\bibinfo {volume} {7}},\ \bibinfo
  {pages} {1479} (\bibinfo {year} {2018})}\BibitemShut {NoStop}%
\bibitem [{\citenamefont {Ruggenthaler}\ \emph {et~al.}(2018)\citenamefont
  {Ruggenthaler}, \citenamefont {Tancogne-Dejean}, \citenamefont {Flick},
  \citenamefont {Appel},\ and\ \citenamefont {Rubio}}]{Ruggenthaler2018}%
  \BibitemOpen
  \bibfield  {author} {\bibinfo {author} {\bibfnamefont {M.}~\bibnamefont
  {Ruggenthaler}}, \bibinfo {author} {\bibfnamefont {N.}~\bibnamefont
  {Tancogne-Dejean}}, \bibinfo {author} {\bibfnamefont {J.}~\bibnamefont
  {Flick}}, \bibinfo {author} {\bibfnamefont {H.}~\bibnamefont {Appel}}, \ and\
  \bibinfo {author} {\bibfnamefont {A.}~\bibnamefont {Rubio}},\ }\href
  {http://dx.doi.org/10.1038/s41570-018-0118} {\bibfield  {journal} {\bibinfo
  {journal} {Nat. Rev. Chem.}\ }\textbf {\bibinfo {volume} {2}},\ \bibinfo
  {pages} {0118} (\bibinfo {year} {2018})}\BibitemShut {NoStop}%
\bibitem [{\citenamefont {Bandrauk}\ \emph {et~al.}(1994)\citenamefont
  {Bandrauk}, \citenamefont {Aubanel},\ and\ \citenamefont
  {Gauthier}}]{Bandrauk3}%
  \BibitemOpen
  \bibfield  {author} {\bibinfo {author} {\bibfnamefont {A.~D.}\ \bibnamefont
  {Bandrauk}}, \bibinfo {author} {\bibfnamefont {E.~E.}\ \bibnamefont
  {Aubanel}}, \ and\ \bibinfo {author} {\bibfnamefont {J.~M.}\ \bibnamefont
  {Gauthier}},\ }\href@noop {} {\emph {\bibinfo {title} {Molecules in Laser
  Fields}}},\ New York,\ Vol.~\bibinfo {volume} {1}\ (\bibinfo  {publisher}
  {Marcel Dekker},\ \bibinfo {year} {1994})\BibitemShut {NoStop}%
\bibitem [{\citenamefont {Hutchison}\ \emph {et~al.}(2012)\citenamefont
  {Hutchison}, \citenamefont {Schwartz}, \citenamefont {Genet}, \citenamefont
  {Devaux},\ and\ \citenamefont {Ebbesen}}]{cavity_exp_Hutchison_AngChem_2012}%
  \BibitemOpen
  \bibfield  {author} {\bibinfo {author} {\bibfnamefont {J.~A.}\ \bibnamefont
  {Hutchison}}, \bibinfo {author} {\bibfnamefont {T.}~\bibnamefont {Schwartz}},
  \bibinfo {author} {\bibfnamefont {C.}~\bibnamefont {Genet}}, \bibinfo
  {author} {\bibfnamefont {E.}~\bibnamefont {Devaux}}, \ and\ \bibinfo {author}
  {\bibfnamefont {T.~W.}\ \bibnamefont {Ebbesen}},\ }\href {\doibase
  10.1002/anie.201107033} {\bibfield  {journal} {\bibinfo  {journal} {Angew.
  Chem. Int. Ed.}\ }\textbf {\bibinfo {volume} {51}},\ \bibinfo {pages} {1592}
  (\bibinfo {year} {2012})}\BibitemShut {NoStop}%
\bibitem [{\citenamefont {Thomas}\ \emph {et~al.}(2016)\citenamefont {Thomas},
  \citenamefont {George}, \citenamefont {Shalabney}, \citenamefont {Dryzhakov},
  \citenamefont {Varma}, \citenamefont {Moran}, \citenamefont {Chervy},
  \citenamefont {Zhong}, \citenamefont {Devaux}, \citenamefont {Genet},
  \citenamefont {Hutchison},\ and\ \citenamefont {Ebbesen}}]{Ebbesen3a}%
  \BibitemOpen
  \bibfield  {author} {\bibinfo {author} {\bibfnamefont {A.}~\bibnamefont
  {Thomas}}, \bibinfo {author} {\bibfnamefont {J.}~\bibnamefont {George}},
  \bibinfo {author} {\bibfnamefont {A.}~\bibnamefont {Shalabney}}, \bibinfo
  {author} {\bibfnamefont {M.}~\bibnamefont {Dryzhakov}}, \bibinfo {author}
  {\bibfnamefont {S.~J.}\ \bibnamefont {Varma}}, \bibinfo {author}
  {\bibfnamefont {J.}~\bibnamefont {Moran}}, \bibinfo {author} {\bibfnamefont
  {T.}~\bibnamefont {Chervy}}, \bibinfo {author} {\bibfnamefont
  {X.}~\bibnamefont {Zhong}}, \bibinfo {author} {\bibfnamefont
  {E.}~\bibnamefont {Devaux}}, \bibinfo {author} {\bibfnamefont
  {C.}~\bibnamefont {Genet}}, \bibinfo {author} {\bibfnamefont {J.~A.}\
  \bibnamefont {Hutchison}}, \ and\ \bibinfo {author} {\bibfnamefont {T.~W.}\
  \bibnamefont {Ebbesen}},\ }\href {\doibase 10.1002/anie.201605504} {\bibfield
   {journal} {\bibinfo  {journal} {Angew. Chem. Int. Ed.}\ }\textbf {\bibinfo
  {volume} {55}},\ \bibinfo {pages} {11462} (\bibinfo {year}
  {2016})}\BibitemShut {NoStop}%
\bibitem [{\citenamefont {Vergauwe}\ \emph {et~al.}(2016)\citenamefont
  {Vergauwe}, \citenamefont {George}, \citenamefont {Chervy}, \citenamefont
  {Hutchison}, \citenamefont {Shalabney}, \citenamefont {Torbeev},\ and\
  \citenamefont {Ebbesen}}]{Ebbesen4a}%
  \BibitemOpen
  \bibfield  {author} {\bibinfo {author} {\bibfnamefont {R.~M.~A.}\
  \bibnamefont {Vergauwe}}, \bibinfo {author} {\bibfnamefont {J.}~\bibnamefont
  {George}}, \bibinfo {author} {\bibfnamefont {T.}~\bibnamefont {Chervy}},
  \bibinfo {author} {\bibfnamefont {J.~A.}\ \bibnamefont {Hutchison}}, \bibinfo
  {author} {\bibfnamefont {A.}~\bibnamefont {Shalabney}}, \bibinfo {author}
  {\bibfnamefont {V.~Y.}\ \bibnamefont {Torbeev}}, \ and\ \bibinfo {author}
  {\bibfnamefont {T.~W.}\ \bibnamefont {Ebbesen}},\ }\href {\doibase
  10.1021/acs.jpclett.6b01869} {\bibfield  {journal} {\bibinfo  {journal} {J.
  Phys. Chem. Lett.}\ }\textbf {\bibinfo {volume} {7}},\ \bibinfo {pages}
  {4159} (\bibinfo {year} {2016})}\BibitemShut {NoStop}%
\bibitem [{\citenamefont {Chervy}\ \emph {et~al.}(2017)\citenamefont {Chervy},
  \citenamefont {Thomas}, \citenamefont {Akiki}, \citenamefont {Vergauwe},
  \citenamefont {Shalabney}, \citenamefont {George}, \citenamefont {Devaux},
  \citenamefont {Hutchison}, \citenamefont {Genet},\ and\ \citenamefont
  {Ebbesen}}]{Ebbesen5a}%
  \BibitemOpen
  \bibfield  {author} {\bibinfo {author} {\bibfnamefont {T.}~\bibnamefont
  {Chervy}}, \bibinfo {author} {\bibfnamefont {A.}~\bibnamefont {Thomas}},
  \bibinfo {author} {\bibfnamefont {E.}~\bibnamefont {Akiki}}, \bibinfo
  {author} {\bibfnamefont {R.~M.~A.}\ \bibnamefont {Vergauwe}}, \bibinfo
  {author} {\bibfnamefont {A.}~\bibnamefont {Shalabney}}, \bibinfo {author}
  {\bibfnamefont {J.}~\bibnamefont {George}}, \bibinfo {author} {\bibfnamefont
  {E.}~\bibnamefont {Devaux}}, \bibinfo {author} {\bibfnamefont {J.~A.}\
  \bibnamefont {Hutchison}}, \bibinfo {author} {\bibfnamefont {C.}~\bibnamefont
  {Genet}}, \ and\ \bibinfo {author} {\bibfnamefont {T.~W.}\ \bibnamefont
  {Ebbesen}},\ }\href {\doibase 10.1021/acsphotonics.7b00677} {\bibfield
  {journal} {\bibinfo  {journal} {{ACS} Photonics}\ }\textbf {\bibinfo {volume}
  {5}},\ \bibinfo {pages} {217} (\bibinfo {year} {2017})}\BibitemShut {NoStop}%
\bibitem [{\citenamefont {Zhong}\ \emph {et~al.}(2016)\citenamefont {Zhong},
  \citenamefont {Chervy}, \citenamefont {Wang}, \citenamefont {George},
  \citenamefont {Thomas}, \citenamefont {Hutchison}, \citenamefont {Devaux},
  \citenamefont {Genet},\ and\ \citenamefont
  {Ebbesen}}]{cavity_Zhong_AngChem_2016}%
  \BibitemOpen
  \bibfield  {author} {\bibinfo {author} {\bibfnamefont {X.}~\bibnamefont
  {Zhong}}, \bibinfo {author} {\bibfnamefont {T.}~\bibnamefont {Chervy}},
  \bibinfo {author} {\bibfnamefont {S.}~\bibnamefont {Wang}}, \bibinfo {author}
  {\bibfnamefont {J.}~\bibnamefont {George}}, \bibinfo {author} {\bibfnamefont
  {A.}~\bibnamefont {Thomas}}, \bibinfo {author} {\bibfnamefont {J.~A.}\
  \bibnamefont {Hutchison}}, \bibinfo {author} {\bibfnamefont {E.}~\bibnamefont
  {Devaux}}, \bibinfo {author} {\bibfnamefont {C.}~\bibnamefont {Genet}}, \
  and\ \bibinfo {author} {\bibfnamefont {T.~W.}\ \bibnamefont {Ebbesen}},\
  }\href {\doibase 10.1002/anie.201600428} {\bibfield  {journal} {\bibinfo
  {journal} {Angew. Chem. Int. Ed.}\ }\textbf {\bibinfo {volume} {55}},\
  \bibinfo {pages} {6202} (\bibinfo {year} {2016})}\BibitemShut {NoStop}%
\bibitem [{\citenamefont {Vergauwe}\ \emph {et~al.}(2019)\citenamefont
  {Vergauwe}, \citenamefont {Thomas}, \citenamefont {Nagarajan}, \citenamefont
  {Shalabney}, \citenamefont {George}, \citenamefont {Chervy}, \citenamefont
  {Seidel}, \citenamefont {Devaux}, \citenamefont {Torbeev},\ and\
  \citenamefont {Ebbesen}}]{Ebbesen7a}%
  \BibitemOpen
  \bibfield  {author} {\bibinfo {author} {\bibfnamefont {R.~M.~A.}\
  \bibnamefont {Vergauwe}}, \bibinfo {author} {\bibfnamefont {A.}~\bibnamefont
  {Thomas}}, \bibinfo {author} {\bibfnamefont {K.}~\bibnamefont {Nagarajan}},
  \bibinfo {author} {\bibfnamefont {A.}~\bibnamefont {Shalabney}}, \bibinfo
  {author} {\bibfnamefont {J.}~\bibnamefont {George}}, \bibinfo {author}
  {\bibfnamefont {T.}~\bibnamefont {Chervy}}, \bibinfo {author} {\bibfnamefont
  {M.}~\bibnamefont {Seidel}}, \bibinfo {author} {\bibfnamefont
  {E.}~\bibnamefont {Devaux}}, \bibinfo {author} {\bibfnamefont
  {V.}~\bibnamefont {Torbeev}}, \ and\ \bibinfo {author} {\bibfnamefont
  {T.~W.}\ \bibnamefont {Ebbesen}},\ }\href {\doibase 10.1002/anie.201908876}
  {\bibfield  {journal} {\bibinfo  {journal} {Angew. Chem. Int. Ed.}\ }\textbf
  {\bibinfo {volume} {58}},\ \bibinfo {pages} {15324} (\bibinfo {year}
  {2019})}\BibitemShut {NoStop}%
\bibitem [{\citenamefont {Schwartz}\ \emph {et~al.}(2011)\citenamefont
  {Schwartz}, \citenamefont {Hutchison}, \citenamefont {Genet},\ and\
  \citenamefont {Ebbesen}}]{Ebbesen8a}%
  \BibitemOpen
  \bibfield  {author} {\bibinfo {author} {\bibfnamefont {T.}~\bibnamefont
  {Schwartz}}, \bibinfo {author} {\bibfnamefont {J.~A.}\ \bibnamefont
  {Hutchison}}, \bibinfo {author} {\bibfnamefont {C.}~\bibnamefont {Genet}}, \
  and\ \bibinfo {author} {\bibfnamefont {T.~W.}\ \bibnamefont {Ebbesen}},\
  }\href {\doibase 10.1103/physrevlett.106.196405} {\bibfield  {journal}
  {\bibinfo  {journal} {Phys. Rev. Lett.}\ }\textbf {\bibinfo {volume} {106}}
  (\bibinfo {year} {2011}),\ 10.1103/physrevlett.106.196405}\BibitemShut
  {NoStop}%
\bibitem [{\citenamefont {Barachati}\ \emph {et~al.}(2017)\citenamefont
  {Barachati}, \citenamefont {Simon}, \citenamefont {Getmanenko}, \citenamefont
  {Barlow}, \citenamefont {Marder},\ and\ \citenamefont
  {K{\'{e}}na-Cohen}}]{Simon-Janos}%
  \BibitemOpen
  \bibfield  {author} {\bibinfo {author} {\bibfnamefont {F.}~\bibnamefont
  {Barachati}}, \bibinfo {author} {\bibfnamefont {J.}~\bibnamefont {Simon}},
  \bibinfo {author} {\bibfnamefont {Y.~A.}\ \bibnamefont {Getmanenko}},
  \bibinfo {author} {\bibfnamefont {S.}~\bibnamefont {Barlow}}, \bibinfo
  {author} {\bibfnamefont {S.~R.}\ \bibnamefont {Marder}}, \ and\ \bibinfo
  {author} {\bibfnamefont {S.}~\bibnamefont {K{\'{e}}na-Cohen}},\ }\href
  {\doibase 10.1021/acsphotonics.7b00305} {\bibfield  {journal} {\bibinfo
  {journal} {{ACS} Photonics}\ }\textbf {\bibinfo {volume} {5}},\ \bibinfo
  {pages} {119} (\bibinfo {year} {2017})}\BibitemShut {NoStop}%
\bibitem [{\citenamefont {Long}\ and\ \citenamefont {Simpkins}(2014)}]{Long1}%
  \BibitemOpen
  \bibfield  {author} {\bibinfo {author} {\bibfnamefont {J.~P.}\ \bibnamefont
  {Long}}\ and\ \bibinfo {author} {\bibfnamefont {B.~S.}\ \bibnamefont
  {Simpkins}},\ }\href {\doibase 10.1021/ph5003347} {\bibfield  {journal}
  {\bibinfo  {journal} {{ACS} Photonics}\ }\textbf {\bibinfo {volume} {2}},\
  \bibinfo {pages} {130} (\bibinfo {year} {2014})}\BibitemShut {NoStop}%
\bibitem [{\citenamefont {Muallem}\ \emph {et~al.}(2016)\citenamefont
  {Muallem}, \citenamefont {Palatnik}, \citenamefont {Nessim},\ and\
  \citenamefont {Tischler}}]{cavity_Muallem_JPCL_2016}%
  \BibitemOpen
  \bibfield  {author} {\bibinfo {author} {\bibfnamefont {M.}~\bibnamefont
  {Muallem}}, \bibinfo {author} {\bibfnamefont {A.}~\bibnamefont {Palatnik}},
  \bibinfo {author} {\bibfnamefont {G.~D.}\ \bibnamefont {Nessim}}, \ and\
  \bibinfo {author} {\bibfnamefont {Y.~R.}\ \bibnamefont {Tischler}},\ }\href
  {\doibase 10.1021/acs.jpclett.6b00617} {\bibfield  {journal} {\bibinfo
  {journal} {J. Phys. Chem. Lett.}\ }\textbf {\bibinfo {volume} {7}},\ \bibinfo
  {pages} {2002} (\bibinfo {year} {2016})}\BibitemShut {NoStop}%
\bibitem [{\citenamefont {Chikkaraddy}\ \emph {et~al.}(2016)\citenamefont
  {Chikkaraddy}, \citenamefont {de~Nijs}, \citenamefont {Benz}, \citenamefont
  {Barrow}, \citenamefont {Scherman}, \citenamefont {Rosta}, \citenamefont
  {Demetriadou}, \citenamefont {Fox}, \citenamefont {Hess},\ and\ \citenamefont
  {Baumberg}}]{Edina1}%
  \BibitemOpen
  \bibfield  {author} {\bibinfo {author} {\bibfnamefont {R.}~\bibnamefont
  {Chikkaraddy}}, \bibinfo {author} {\bibfnamefont {B.}~\bibnamefont
  {de~Nijs}}, \bibinfo {author} {\bibfnamefont {F.}~\bibnamefont {Benz}},
  \bibinfo {author} {\bibfnamefont {S.~J.}\ \bibnamefont {Barrow}}, \bibinfo
  {author} {\bibfnamefont {O.~A.}\ \bibnamefont {Scherman}}, \bibinfo {author}
  {\bibfnamefont {E.}~\bibnamefont {Rosta}}, \bibinfo {author} {\bibfnamefont
  {A.}~\bibnamefont {Demetriadou}}, \bibinfo {author} {\bibfnamefont
  {P.}~\bibnamefont {Fox}}, \bibinfo {author} {\bibfnamefont {O.}~\bibnamefont
  {Hess}}, \ and\ \bibinfo {author} {\bibfnamefont {J.~J.}\ \bibnamefont
  {Baumberg}},\ }\href {http://dx.doi.org/10.1038/nature17974} {\bibfield
  {journal} {\bibinfo  {journal} {Nature}\ }\textbf {\bibinfo {volume} {535}},\
  \bibinfo {pages} {127} (\bibinfo {year} {2016})}\BibitemShut {NoStop}%
\bibitem [{\citenamefont {Damari}\ \emph {et~al.}(2019)\citenamefont {Damari},
  \citenamefont {Weinberg}, \citenamefont {Krotkov}, \citenamefont {Demina},
  \citenamefont {Akulov}, \citenamefont {Golombek}, \citenamefont {Schwartz},\
  and\ \citenamefont {Fleischer}}]{Fleischer1}%
  \BibitemOpen
  \bibfield  {author} {\bibinfo {author} {\bibfnamefont {R.}~\bibnamefont
  {Damari}}, \bibinfo {author} {\bibfnamefont {O.}~\bibnamefont {Weinberg}},
  \bibinfo {author} {\bibfnamefont {D.}~\bibnamefont {Krotkov}}, \bibinfo
  {author} {\bibfnamefont {N.}~\bibnamefont {Demina}}, \bibinfo {author}
  {\bibfnamefont {K.}~\bibnamefont {Akulov}}, \bibinfo {author} {\bibfnamefont
  {A.}~\bibnamefont {Golombek}}, \bibinfo {author} {\bibfnamefont
  {T.}~\bibnamefont {Schwartz}}, \ and\ \bibinfo {author} {\bibfnamefont
  {S.}~\bibnamefont {Fleischer}},\ }\href {\doibase 10.1038/s41467-019-11130-y}
  {\bibfield  {journal} {\bibinfo  {journal} {Nature Communications}\ }\textbf
  {\bibinfo {volume} {10}} (\bibinfo {year} {2019}),\
  10.1038/s41467-019-11130-y}\BibitemShut {NoStop}%
\bibitem [{\citenamefont {Galego}\ \emph {et~al.}(2015)\citenamefont {Galego},
  \citenamefont {Garcia-Vidal},\ and\ \citenamefont
  {Feist}}]{cavity_Galego_PRX_2015}%
  \BibitemOpen
  \bibfield  {author} {\bibinfo {author} {\bibfnamefont {J.}~\bibnamefont
  {Galego}}, \bibinfo {author} {\bibfnamefont {F.~J.}\ \bibnamefont
  {Garcia-Vidal}}, \ and\ \bibinfo {author} {\bibfnamefont {J.}~\bibnamefont
  {Feist}},\ }\href {\doibase 10.1103/PhysRevX.5.041022} {\bibfield  {journal}
  {\bibinfo  {journal} {Phys. Rev. X}\ }\textbf {\bibinfo {volume} {5}},\
  \bibinfo {pages} {1} (\bibinfo {year} {2015})}\BibitemShut {NoStop}%
\bibitem [{\citenamefont {Galego}\ \emph {et~al.}(2016)\citenamefont {Galego},
  \citenamefont {Garcia-Vidal},\ and\ \citenamefont
  {Feist}}]{cavity_Galego_NatCommun_2016}%
  \BibitemOpen
  \bibfield  {author} {\bibinfo {author} {\bibfnamefont {J.}~\bibnamefont
  {Galego}}, \bibinfo {author} {\bibfnamefont {F.~J.}\ \bibnamefont
  {Garcia-Vidal}}, \ and\ \bibinfo {author} {\bibfnamefont {J.}~\bibnamefont
  {Feist}},\ }\href {\doibase 10.1038/ncomms13841} {\bibfield  {journal}
  {\bibinfo  {journal} {Nat. Commun.}\ }\textbf {\bibinfo {volume} {7}},\
  \bibinfo {pages} {13841} (\bibinfo {year} {2016})}\BibitemShut {NoStop}%
\bibitem [{\citenamefont {Kowalewski}\ \emph
  {et~al.}(2016{\natexlab{a}})\citenamefont {Kowalewski}, \citenamefont
  {Bennett},\ and\ \citenamefont {Mukamel}}]{markus1a}%
  \BibitemOpen
  \bibfield  {author} {\bibinfo {author} {\bibfnamefont {M.}~\bibnamefont
  {Kowalewski}}, \bibinfo {author} {\bibfnamefont {K.}~\bibnamefont {Bennett}},
  \ and\ \bibinfo {author} {\bibfnamefont {S.}~\bibnamefont {Mukamel}},\ }\href
  {\doibase 10.1063/1.4941053} {\bibfield  {journal} {\bibinfo  {journal} {J.
  Chem. Phys.}\ }\textbf {\bibinfo {volume} {144}},\ \bibinfo {pages} {054309}
  (\bibinfo {year} {2016}{\natexlab{a}})}\BibitemShut {NoStop}%
\bibitem [{\citenamefont {Kowalewski}\ \emph
  {et~al.}(2016{\natexlab{b}})\citenamefont {Kowalewski}, \citenamefont
  {Bennett},\ and\ \citenamefont {Mukamel}}]{cavity_Kowalewski_JPCL_2016}%
  \BibitemOpen
  \bibfield  {author} {\bibinfo {author} {\bibfnamefont {M.}~\bibnamefont
  {Kowalewski}}, \bibinfo {author} {\bibfnamefont {K.}~\bibnamefont {Bennett}},
  \ and\ \bibinfo {author} {\bibfnamefont {S.}~\bibnamefont {Mukamel}},\ }\href
  {\doibase 10.1021/acs.jpclett.6b00864} {\bibfield  {journal} {\bibinfo
  {journal} {J. Phys. Chem. Lett.}\ }\textbf {\bibinfo {volume} {7}},\ \bibinfo
  {pages} {2050} (\bibinfo {year} {2016}{\natexlab{b}})}\BibitemShut {NoStop}%
\bibitem [{\citenamefont {Luk}\ \emph {et~al.}(2017)\citenamefont {Luk},
  \citenamefont {Feist}, \citenamefont {Toppari},\ and\ \citenamefont
  {Groenhof}}]{cavity_Luk_JCTC_2017}%
  \BibitemOpen
  \bibfield  {author} {\bibinfo {author} {\bibfnamefont {H.~L.}\ \bibnamefont
  {Luk}}, \bibinfo {author} {\bibfnamefont {J.}~\bibnamefont {Feist}}, \bibinfo
  {author} {\bibfnamefont {J.~J.}\ \bibnamefont {Toppari}}, \ and\ \bibinfo
  {author} {\bibfnamefont {G.}~\bibnamefont {Groenhof}},\ }\href {\doibase
  10.1021/acs.jctc.7b00388} {\bibfield  {journal} {\bibinfo  {journal} {J.
  Chem. Theory Comput.}\ }\textbf {\bibinfo {volume} {13}},\ \bibinfo {pages}
  {4324} (\bibinfo {year} {2017})}\BibitemShut {NoStop}%
\bibitem [{\citenamefont {Groenhof}\ and\ \citenamefont
  {Toppari}(2018)}]{gerit2a}%
  \BibitemOpen
  \bibfield  {author} {\bibinfo {author} {\bibfnamefont {G.}~\bibnamefont
  {Groenhof}}\ and\ \bibinfo {author} {\bibfnamefont {J.~J.}\ \bibnamefont
  {Toppari}},\ }\href {\doibase 10.1021/acs.jpclett.8b02032} {\bibfield
  {journal} {\bibinfo  {journal} {J. Phys. Chem. Lett.}\ }\textbf {\bibinfo
  {volume} {9}},\ \bibinfo {pages} {4848} (\bibinfo {year} {2018})}\BibitemShut
  {NoStop}%
\bibitem [{\citenamefont {Groenhof}\ \emph {et~al.}(2019)\citenamefont
  {Groenhof}, \citenamefont {Climent}, \citenamefont {Feist}, \citenamefont
  {Morozov},\ and\ \citenamefont {Toppari}}]{gerit3a}%
  \BibitemOpen
  \bibfield  {author} {\bibinfo {author} {\bibfnamefont {G.}~\bibnamefont
  {Groenhof}}, \bibinfo {author} {\bibfnamefont {C.}~\bibnamefont {Climent}},
  \bibinfo {author} {\bibfnamefont {J.}~\bibnamefont {Feist}}, \bibinfo
  {author} {\bibfnamefont {D.}~\bibnamefont {Morozov}}, \ and\ \bibinfo
  {author} {\bibfnamefont {J.~J.}\ \bibnamefont {Toppari}},\ }\href {\doibase
  10.1021/acs.jpclett.9b02192} {\bibfield  {journal} {\bibinfo  {journal} {J.
  Phys. Chem. Lett.}\ }\textbf {\bibinfo {volume} {10}},\ \bibinfo {pages}
  {5476} (\bibinfo {year} {2019})}\BibitemShut {NoStop}%
\bibitem [{\citenamefont {Yuen-Zhou}\ \emph {et~al.}(2018)\citenamefont
  {Yuen-Zhou}, \citenamefont {Saikin},\ and\ \citenamefont {Menon}}]{joel2a}%
  \BibitemOpen
  \bibfield  {author} {\bibinfo {author} {\bibfnamefont {J.}~\bibnamefont
  {Yuen-Zhou}}, \bibinfo {author} {\bibfnamefont {S.~K.}\ \bibnamefont
  {Saikin}}, \ and\ \bibinfo {author} {\bibfnamefont {V.~M.}\ \bibnamefont
  {Menon}},\ }\href {\doibase 10.1021/acs.jpclett.8b02980} {\bibfield
  {journal} {\bibinfo  {journal} {J. Phys. Chem. Lett.}\ }\textbf {\bibinfo
  {volume} {9}},\ \bibinfo {pages} {6511} (\bibinfo {year} {2018})}\BibitemShut
  {NoStop}%
\bibitem [{\citenamefont {K{\'{e}}na-Cohen}\ and\ \citenamefont
  {Yuen-Zhou}(2019)}]{joel3a}%
  \BibitemOpen
  \bibfield  {author} {\bibinfo {author} {\bibfnamefont {S.}~\bibnamefont
  {K{\'{e}}na-Cohen}}\ and\ \bibinfo {author} {\bibfnamefont {J.}~\bibnamefont
  {Yuen-Zhou}},\ }\href {\doibase 10.1021/acscentsci.9b00219} {\bibfield
  {journal} {\bibinfo  {journal} {{ACS} Central Science}\ }\textbf {\bibinfo
  {volume} {5}},\ \bibinfo {pages} {386} (\bibinfo {year} {2019})}\BibitemShut
  {NoStop}%
\bibitem [{\citenamefont {Campos-Gonzalez-Angulo}\ \emph
  {et~al.}(2019)\citenamefont {Campos-Gonzalez-Angulo}, \citenamefont
  {Ribeiro},\ and\ \citenamefont {Yuen-Zhou}}]{joel4a}%
  \BibitemOpen
  \bibfield  {author} {\bibinfo {author} {\bibfnamefont {J.~A.}\ \bibnamefont
  {Campos-Gonzalez-Angulo}}, \bibinfo {author} {\bibfnamefont {R.~F.}\
  \bibnamefont {Ribeiro}}, \ and\ \bibinfo {author} {\bibfnamefont
  {J.}~\bibnamefont {Yuen-Zhou}},\ }\href {\doibase 10.1038/s41467-019-12636-1}
  {\bibfield  {journal} {\bibinfo  {journal} {Nat. Commun.}\ }\textbf {\bibinfo
  {volume} {10}} (\bibinfo {year} {2019}),\
  10.1038/s41467-019-12636-1}\BibitemShut {NoStop}%
\bibitem [{\citenamefont {Herrera}\ and\ \citenamefont
  {Spano}(2016)}]{cavity_Herrera_PRL_2016}%
  \BibitemOpen
  \bibfield  {author} {\bibinfo {author} {\bibfnamefont {F.}~\bibnamefont
  {Herrera}}\ and\ \bibinfo {author} {\bibfnamefont {F.~C.}\ \bibnamefont
  {Spano}},\ }\href {\doibase 10.1103/PhysRevLett.116.238301} {\bibfield
  {journal} {\bibinfo  {journal} {Phys. Rev. Lett.}\ }\textbf {\bibinfo
  {volume} {116}},\ \bibinfo {pages} {238301} (\bibinfo {year}
  {2016})}\BibitemShut {NoStop}%
\bibitem [{\citenamefont {Herrera}\ and\ \citenamefont
  {Spano}(2017)}]{dark_vibronic_polaritons_Herrera_PRL_2017}%
  \BibitemOpen
  \bibfield  {author} {\bibinfo {author} {\bibfnamefont {F.}~\bibnamefont
  {Herrera}}\ and\ \bibinfo {author} {\bibfnamefont {F.~C.}\ \bibnamefont
  {Spano}},\ }\href {\doibase 10.1103/PhysRevLett.118.223601} {\bibfield
  {journal} {\bibinfo  {journal} {Phys. Rev. Lett.}\ }\textbf {\bibinfo
  {volume} {118}},\ \bibinfo {pages} {223601} (\bibinfo {year}
  {2017})}\BibitemShut {NoStop}%
\bibitem [{\citenamefont {Flick}\ \emph
  {et~al.}(2017{\natexlab{a}})\citenamefont {Flick}, \citenamefont
  {Ruggenthaler}, \citenamefont {Appel},\ and\ \citenamefont
  {Rubio}}]{cavity_Flick_PNAS_2017}%
  \BibitemOpen
  \bibfield  {author} {\bibinfo {author} {\bibfnamefont {J.}~\bibnamefont
  {Flick}}, \bibinfo {author} {\bibfnamefont {M.}~\bibnamefont {Ruggenthaler}},
  \bibinfo {author} {\bibfnamefont {H.}~\bibnamefont {Appel}}, \ and\ \bibinfo
  {author} {\bibfnamefont {A.}~\bibnamefont {Rubio}},\ }\href {\doibase
  10.1073/pnas.1615509114} {\bibfield  {journal} {\bibinfo  {journal} {Proc.
  Natl. Acad. Sci.}\ }\textbf {\bibinfo {volume} {114}},\ \bibinfo {pages}
  {3026} (\bibinfo {year} {2017}{\natexlab{a}})}\BibitemShut {NoStop}%
\bibitem [{\citenamefont {Flick}\ \emph
  {et~al.}(2017{\natexlab{b}})\citenamefont {Flick}, \citenamefont {Appel},
  \citenamefont {Ruggenthaler},\ and\ \citenamefont
  {Rubio}}]{cavityBO_Flick_JCTC_2017}%
  \BibitemOpen
  \bibfield  {author} {\bibinfo {author} {\bibfnamefont {J.}~\bibnamefont
  {Flick}}, \bibinfo {author} {\bibfnamefont {H.}~\bibnamefont {Appel}},
  \bibinfo {author} {\bibfnamefont {M.}~\bibnamefont {Ruggenthaler}}, \ and\
  \bibinfo {author} {\bibfnamefont {A.}~\bibnamefont {Rubio}},\ }\href
  {\doibase 10.1021/acs.jctc.6b01126} {\bibfield  {journal} {\bibinfo
  {journal} {J. Chem. Theory Comput.}\ }\textbf {\bibinfo {volume} {13}},\
  \bibinfo {pages} {1616} (\bibinfo {year} {2017}{\natexlab{b}})}\BibitemShut
  {NoStop}%
\bibitem [{\citenamefont
  {Vendrell}(2018{\natexlab{a}})}]{cavity_MCTDH_Vendrell_CP_2018}%
  \BibitemOpen
  \bibfield  {author} {\bibinfo {author} {\bibfnamefont {O.}~\bibnamefont
  {Vendrell}},\ }\href {\doibase
  https://doi.org/10.1016/j.chemphys.2018.02.008} {\bibfield  {journal}
  {\bibinfo  {journal} {Chem. Phys.}\ }\textbf {\bibinfo {volume} {509}},\
  \bibinfo {pages} {55 } (\bibinfo {year} {2018}{\natexlab{a}})}\BibitemShut
  {NoStop}%
\bibitem [{\citenamefont {Vendrell}(2018{\natexlab{b}})}]{Oriol2a}%
  \BibitemOpen
  \bibfield  {author} {\bibinfo {author} {\bibfnamefont {O.}~\bibnamefont
  {Vendrell}},\ }\href {\doibase 10.1103/physrevlett.121.253001} {\bibfield
  {journal} {\bibinfo  {journal} {Phys. Rev. Lett.}\ }\textbf {\bibinfo
  {volume} {121}} (\bibinfo {year} {2018}{\natexlab{b}}),\
  10.1103/physrevlett.121.253001}\BibitemShut {NoStop}%
\bibitem [{\citenamefont {Fregoni}\ \emph {et~al.}(2019)\citenamefont
  {Fregoni}, \citenamefont {Granucci}, \citenamefont {Persico},\ and\
  \citenamefont {Corni}}]{Fregoni2019}%
  \BibitemOpen
  \bibfield  {author} {\bibinfo {author} {\bibfnamefont {J.}~\bibnamefont
  {Fregoni}}, \bibinfo {author} {\bibfnamefont {G.}~\bibnamefont {Granucci}},
  \bibinfo {author} {\bibfnamefont {M.}~\bibnamefont {Persico}}, \ and\
  \bibinfo {author} {\bibfnamefont {S.}~\bibnamefont {Corni}},\ }\href
  {\doibase 10.1016/j.chempr.2019.11.001} {\bibfield  {journal} {\bibinfo
  {journal} {Chem}\ } (\bibinfo {year} {2019}),\
  10.1016/j.chempr.2019.11.001}\BibitemShut {NoStop}%
\bibitem [{\citenamefont {Sch\"{a}fer}\ \emph {et~al.}(2019)\citenamefont
  {Sch\"{a}fer}, \citenamefont {Ruggenthaler}, \citenamefont {Appel},\ and\
  \citenamefont {Rubio}}]{Schfer2019}%
  \BibitemOpen
  \bibfield  {author} {\bibinfo {author} {\bibfnamefont {C.}~\bibnamefont
  {Sch\"{a}fer}}, \bibinfo {author} {\bibfnamefont {M.}~\bibnamefont
  {Ruggenthaler}}, \bibinfo {author} {\bibfnamefont {H.}~\bibnamefont {Appel}},
  \ and\ \bibinfo {author} {\bibfnamefont {A.}~\bibnamefont {Rubio}},\ }\href
  {\doibase 10.1073/pnas.1814178116} {\bibfield  {journal} {\bibinfo  {journal}
  {Proc. Natl. Acad. Sci.}\ }\textbf {\bibinfo {volume} {116}},\ \bibinfo
  {pages} {4883} (\bibinfo {year} {2019})}\BibitemShut {NoStop}%
\bibitem [{\citenamefont {Szidarovszky}\ \emph
  {et~al.}(2018{\natexlab{a}})\citenamefont {Szidarovszky}, \citenamefont
  {Hal{\'{a}}sz}, \citenamefont {Cs{\'{a}}sz{\'{a}}r}, \citenamefont
  {Cederbaum},\ and\ \citenamefont {Vib{\'{o}}k}}]{Tamas1a}%
  \BibitemOpen
  \bibfield  {author} {\bibinfo {author} {\bibfnamefont {T.}~\bibnamefont
  {Szidarovszky}}, \bibinfo {author} {\bibfnamefont {G.~J.}\ \bibnamefont
  {Hal{\'{a}}sz}}, \bibinfo {author} {\bibfnamefont {A.~G.}\ \bibnamefont
  {Cs{\'{a}}sz{\'{a}}r}}, \bibinfo {author} {\bibfnamefont {L.~S.}\
  \bibnamefont {Cederbaum}}, \ and\ \bibinfo {author} {\bibfnamefont
  {{\'{A}}.}~\bibnamefont {Vib{\'{o}}k}},\ }\href {\doibase
  10.1021/acs.jpclett.8b02609} {\bibfield  {journal} {\bibinfo  {journal} {J.
  Phys. Chem. Lett.}\ }\textbf {\bibinfo {volume} {9}},\ \bibinfo {pages}
  {6215} (\bibinfo {year} {2018}{\natexlab{a}})}\BibitemShut {NoStop}%
\bibitem [{\citenamefont {Csehi}\ \emph {et~al.}(2017)\citenamefont {Csehi},
  \citenamefont {Hal\'asz}, \citenamefont {Cederbaum},\ and\ \citenamefont
  {Vib\'ok}}]{Gabi6}%
  \BibitemOpen
  \bibfield  {author} {\bibinfo {author} {\bibfnamefont {A.}~\bibnamefont
  {Csehi}}, \bibinfo {author} {\bibfnamefont {G.~J.}\ \bibnamefont {Hal\'asz}},
  \bibinfo {author} {\bibfnamefont {L.~S.}\ \bibnamefont {Cederbaum}}, \ and\
  \bibinfo {author} {\bibfnamefont {A.}~\bibnamefont {Vib\'ok}},\ }\href
  {\doibase 10.1021/acs.jpclett.7b00413} {\bibfield  {journal} {\bibinfo
  {journal} {J. Phys. Chem. Lett.}\ }\textbf {\bibinfo {volume} {8}},\ \bibinfo
  {pages} {1624} (\bibinfo {year} {2017})}\BibitemShut {NoStop}%
\bibitem [{\citenamefont {Csehi}\ \emph {et~al.}(2019)\citenamefont {Csehi},
  \citenamefont {Kowalewski}, \citenamefont {Hal{\'{a}}sz},\ and\ \citenamefont
  {Vib{\'{o}}k}}]{Agi2}%
  \BibitemOpen
  \bibfield  {author} {\bibinfo {author} {\bibfnamefont {A.}~\bibnamefont
  {Csehi}}, \bibinfo {author} {\bibfnamefont {M.}~\bibnamefont {Kowalewski}},
  \bibinfo {author} {\bibfnamefont {G.~J.}\ \bibnamefont {Hal{\'{a}}sz}}, \
  and\ \bibinfo {author} {\bibfnamefont {{\'{A}}.}~\bibnamefont
  {Vib{\'{o}}k}},\ }\href {\doibase 10.1088/1367-2630/ab3fcc} {\bibfield
  {journal} {\bibinfo  {journal} {New J. Phys.}\ }\textbf {\bibinfo {volume}
  {21}},\ \bibinfo {pages} {093040} (\bibinfo {year} {2019})}\BibitemShut
  {NoStop}%
\bibitem [{\citenamefont {Csehi}\ \emph {et~al.}(2018)\citenamefont {Csehi},
  \citenamefont {Hal{\'{a}}sz},\ and\ \citenamefont {Vib{\'{o}}k}}]{Agi3}%
  \BibitemOpen
  \bibfield  {author} {\bibinfo {author} {\bibfnamefont {A.}~\bibnamefont
  {Csehi}}, \bibinfo {author} {\bibfnamefont {G.~J.}\ \bibnamefont
  {Hal{\'{a}}sz}}, \ and\ \bibinfo {author} {\bibfnamefont
  {{\'{A}}.}~\bibnamefont {Vib{\'{o}}k}},\ }\href {\doibase
  10.1016/j.chemphys.2017.12.017} {\bibfield  {journal} {\bibinfo  {journal}
  {Chem. Phys.}\ }\textbf {\bibinfo {volume} {509}},\ \bibinfo {pages} {91}
  (\bibinfo {year} {2018})}\BibitemShut {NoStop}%
\bibitem [{\citenamefont {Cohen-Tannoudji}\ \emph {et~al.}(2004)\citenamefont
  {Cohen-Tannoudji}, \citenamefont {Dupont-Roc},\ and\ \citenamefont
  {Grynberg}}]{Cohen-Tannoudji}%
  \BibitemOpen
  \bibfield  {author} {\bibinfo {author} {\bibfnamefont {C.}~\bibnamefont
  {Cohen-Tannoudji}}, \bibinfo {author} {\bibfnamefont {J.}~\bibnamefont
  {Dupont-Roc}}, \ and\ \bibinfo {author} {\bibfnamefont {G.}~\bibnamefont
  {Grynberg}},\ }\href {\doibase 10.1002/9783527617197} {\emph {\bibinfo
  {title} {Atom-Photon Interactions: Basic Processes and Applications}}},\
  Weinheim\ (\bibinfo  {publisher} {Wiley-VCH Verlag GmbH and Co. KGaA},\
  \bibinfo {year} {2004})\BibitemShut {NoStop}%
\bibitem [{\citenamefont {Magnier}\ \emph {et~al.}(1993)\citenamefont
  {Magnier}, \citenamefont {Milli\'e}, \citenamefont {Dulieu},\ and\
  \citenamefont {Masnou-Seeuws}}]{Na2_PEC_Magnier_JCP_1993}%
  \BibitemOpen
  \bibfield  {author} {\bibinfo {author} {\bibfnamefont {S.}~\bibnamefont
  {Magnier}}, \bibinfo {author} {\bibfnamefont {P.}~\bibnamefont {Milli\'e}},
  \bibinfo {author} {\bibfnamefont {O.}~\bibnamefont {Dulieu}}, \ and\ \bibinfo
  {author} {\bibfnamefont {F.}~\bibnamefont {Masnou-Seeuws}},\ }\href@noop {}
  {\bibfield  {journal} {\bibinfo  {journal} {J. Chem. Phys.}\ }\textbf
  {\bibinfo {volume} {98}},\ \bibinfo {pages} {7113} (\bibinfo {year}
  {1993})}\BibitemShut {NoStop}%
\bibitem [{\citenamefont {Zemke}\ \emph {et~al.}(1981)\citenamefont {Zemke},
  \citenamefont {Verma}, \citenamefont {Vu},\ and\ \citenamefont
  {Stwalley}}]{Na2_TDM_Zemke_JMS_1981}%
  \BibitemOpen
  \bibfield  {author} {\bibinfo {author} {\bibfnamefont {W.~T.}\ \bibnamefont
  {Zemke}}, \bibinfo {author} {\bibfnamefont {K.~K.}\ \bibnamefont {Verma}},
  \bibinfo {author} {\bibfnamefont {T.}~\bibnamefont {Vu}}, \ and\ \bibinfo
  {author} {\bibfnamefont {W.~C.}\ \bibnamefont {Stwalley}},\ }\href@noop {}
  {\bibfield  {journal} {\bibinfo  {journal} {J. Mol. Spectrosc.}\ }\textbf
  {\bibinfo {volume} {85}},\ \bibinfo {pages} {150} (\bibinfo {year}
  {1981})}\BibitemShut {NoStop}%
\bibitem [{\citenamefont {Houdr{\'{e}}}\ \emph {et~al.}(1996)\citenamefont
  {Houdr{\'{e}}}, \citenamefont {Stanley},\ and\ \citenamefont
  {Ilegems}}]{Houdr1996}%
  \BibitemOpen
  \bibfield  {author} {\bibinfo {author} {\bibfnamefont {R.}~\bibnamefont
  {Houdr{\'{e}}}}, \bibinfo {author} {\bibfnamefont {R.~P.}\ \bibnamefont
  {Stanley}}, \ and\ \bibinfo {author} {\bibfnamefont {M.}~\bibnamefont
  {Ilegems}},\ }\href {\doibase 10.1103/physreva.53.2711} {\bibfield  {journal}
  {\bibinfo  {journal} {Phys. Rev. A}\ }\textbf {\bibinfo {volume} {53}},\
  \bibinfo {pages} {2711} (\bibinfo {year} {1996})}\BibitemShut {NoStop}%
\bibitem [{\citenamefont {Szidarovszky}\ \emph
  {et~al.}(2018{\natexlab{b}})\citenamefont {Szidarovszky}, \citenamefont
  {Hal{\'{a}}sz}, \citenamefont {Cs{\'{a}}sz{\'{a}}r}, \citenamefont
  {Cederbaum},\ and\ \citenamefont
  {Vib{\'{o}}k}}]{LICI_in_spectrum_Szidarovszky_JPCL_2018}%
  \BibitemOpen
  \bibfield  {author} {\bibinfo {author} {\bibfnamefont {T.}~\bibnamefont
  {Szidarovszky}}, \bibinfo {author} {\bibfnamefont {G.~J.}\ \bibnamefont
  {Hal{\'{a}}sz}}, \bibinfo {author} {\bibfnamefont {A.~G.}\ \bibnamefont
  {Cs{\'{a}}sz{\'{a}}r}}, \bibinfo {author} {\bibfnamefont {L.~S.}\
  \bibnamefont {Cederbaum}}, \ and\ \bibinfo {author} {\bibfnamefont
  {{\'{A}}.}~\bibnamefont {Vib{\'{o}}k}},\ }\href {\doibase
  10.1021/acs.jpclett.8b01102} {\bibfield  {journal} {\bibinfo  {journal} {J.
  Phys. Chem. Lett.}\ }\textbf {\bibinfo {volume} {9}},\ \bibinfo {pages}
  {2739} (\bibinfo {year} {2018}{\natexlab{b}})}\BibitemShut {NoStop}%
\bibitem [{\citenamefont {Bunker}\ and\ \citenamefont
  {Jensen}(1998)}]{BunkerJensen}%
  \BibitemOpen
  \bibfield  {author} {\bibinfo {author} {\bibfnamefont {P.~R.}\ \bibnamefont
  {Bunker}}\ and\ \bibinfo {author} {\bibfnamefont {P.}~\bibnamefont
  {Jensen}},\ }\href@noop {} {\emph {\bibinfo {title} {Molecular Symmetry and
  Spectroscopy}}},\ Ottawa\ (\bibinfo  {publisher} {NRC Research Press},\
  \bibinfo {year} {1998})\BibitemShut {NoStop}%
\bibitem [{\citenamefont {Szidarovszky}\ \emph {et~al.}(2010)\citenamefont
  {Szidarovszky}, \citenamefont {Cs{\'{a}}sz{\'{a}}r},\ and\ \citenamefont
  {Czak{\'{o}}}}]{D2FOPI_Szidarovszky_PCCP_2010}%
  \BibitemOpen
  \bibfield  {author} {\bibinfo {author} {\bibfnamefont {T.}~\bibnamefont
  {Szidarovszky}}, \bibinfo {author} {\bibfnamefont {A.~G.}\ \bibnamefont
  {Cs{\'{a}}sz{\'{a}}r}}, \ and\ \bibinfo {author} {\bibfnamefont
  {G.}~\bibnamefont {Czak{\'{o}}}},\ }\href {\doibase 10.1039/c001124j}
  {\bibfield  {journal} {\bibinfo  {journal} {Phys. Chem. Chem. Phys.}\
  }\textbf {\bibinfo {volume} {12}},\ \bibinfo {pages} {8373} (\bibinfo {year}
  {2010})}\BibitemShut {NoStop}%
\bibitem [{\citenamefont {Jaynes}\ and\ \citenamefont
  {Cummings}(1963)}]{Jaynes_Cummings_1963}%
  \BibitemOpen
  \bibfield  {author} {\bibinfo {author} {\bibfnamefont {E.~T.}\ \bibnamefont
  {Jaynes}}\ and\ \bibinfo {author} {\bibfnamefont {F.~W.}\ \bibnamefont
  {Cummings}},\ }\href {\doibase 10.1109/PROC.1963.1664} {\bibfield  {journal}
  {\bibinfo  {journal} {Proc. IEEE}\ }\textbf {\bibinfo {volume} {51}},\
  \bibinfo {pages} {89} (\bibinfo {year} {1963})}\BibitemShut {NoStop}%
\bibitem [{\citenamefont {Autler}\ and\ \citenamefont
  {Townes}(1955)}]{AutlerTownes_original}%
  \BibitemOpen
  \bibfield  {author} {\bibinfo {author} {\bibfnamefont {S.~H.}\ \bibnamefont
  {Autler}}\ and\ \bibinfo {author} {\bibfnamefont {C.~H.}\ \bibnamefont
  {Townes}},\ }\href@noop {} {\bibfield  {journal} {\bibinfo  {journal} {Phys.
  Rev.}\ }\textbf {\bibinfo {volume} {100}},\ \bibinfo {pages} {703} (\bibinfo
  {year} {1955})}\BibitemShut {NoStop}%
\bibitem [{\citenamefont {K\"oppel}\ \emph {et~al.}(1984)\citenamefont
  {K\"oppel}, \citenamefont {Domcke},\ and\ \citenamefont
  {Cederbaum}}]{Cederbaum_multimode}%
  \BibitemOpen
  \bibfield  {author} {\bibinfo {author} {\bibfnamefont {H.}~\bibnamefont
  {K\"oppel}}, \bibinfo {author} {\bibfnamefont {W.}~\bibnamefont {Domcke}}, \
  and\ \bibinfo {author} {\bibfnamefont {L.~S.}\ \bibnamefont {Cederbaum}},\
  }\href@noop {} {\bibfield  {journal} {\bibinfo  {journal} {Adv. Chem. Phys.}\
  }\textbf {\bibinfo {volume} {57}},\ \bibinfo {pages} {59} (\bibinfo {year}
  {1984})}\BibitemShut {NoStop}%
\end{thebibliography}%

\end{document}